\pdfoutput=1

\documentclass[onecolumn]{aastex6}
\usepackage{amsmath}
\usepackage{graphicx}
\usepackage{float}

\newcommand{\sn}[2]{#1 \times 10^{#2}}
\newcommand{\integ}[4]{\int_{#1}^{#2} \! {#3} \, \mathrm{d} {#4}}

\shorttitle{Line Broadening in Flares}
\shortauthors{Kowalski, et al.}

\begin{document}

\title{Hydrogen Balmer Line Broadening in Solar and Stellar Flares}
\author{Adam F. Kowalski}
\affil{Department of Astrophysical and Planetary Sciences, University of Colorado Boulder, 2000 Colorado Ave, Boulder, CO 80305, USA.}
\affil{National Solar Observatory, University of Colorado Boulder, 3665 Discovery Drive, Boulder, CO 80303, USA.}
\email{Adam.Kowalski@lasp.colorado.edu}
\author{Joel C. Allred}
\affil{NASA/Goddard Space Flight Center, Code 671, Greenbelt, MD 20771}
\author{Han Uitenbroek}
\affil{National Solar Observatory, University of Colorado Boulder, 3665 Discovery Drive, Boulder, CO 80303, USA.}
\author{Pier-Emmanuel Tremblay}
\affil{Department of Physics, University of Warwick, Coventry CV47AL, UK.}
\author{Stephen Brown}
\affil{School of Physics and Astronomy, Kelvin Building, University of Glasgow, G12 8QQ, Scotland.}
\author{Mats Carlsson}
\affil{Institute of Theoretical Astrophysics, University of Oslo, PO Box 1029 Blindern, 0315 Oslo, Norway. }
\author{Rachel A. Osten}
\affil{Space Telescope Science Institute, 3700 San Martin Drive, Baltimore, MD 21218, USA.}
\author{John P. Wisniewski}
\affil{Homer L. Dodge Department of Physics and Astronomy, University of Oklahoma, 440 W. Brooks Street, Norman, OK 73019, USA.}
\author{Suzanne L. Hawley}
\affil{University of Washington Department of Astronomy, 3910 15th Ave
  NE, Seattle, WA 98195, USA.}

\begin{abstract}

The broadening of the hydrogen lines during flares is thought to result from increased charge (electron, proton) density in the flare chromosphere.  However, disagreements between theory and modeling prescriptions have precluded an accurate diagnostic of the degree of ionization and compression resulting from flare heating in the chromosphere.  To resolve this issue,  we have incorporated the unified theory of electric pressure broadening of the hydrogen lines into the non-LTE radiative transfer code RH.  This broadening prescription produces a much more realistic spectrum of the quiescent, A0 star Vega compared to the analytic approximations used as a damping parameter in the Voigt profiles.   We test recent radiative-hydrodynamic (RHD) simulations of the atmospheric response to high nonthermal electron beam fluxes with the new broadening prescription and find that the Balmer lines are overbroadened at the densest times in the simulations.  Adding many simultaneously heated and cooling model loops as a ``multithread'' model improves the agreement with the observations.  We revisit the three-component phenomenological flare model of the YZ CMi Megaflare using recent and new RHD models.  The evolution of the broadening, line flux ratios, and continuum flux ratios are well-reproduced by a multithread model with high-flux nonthermal electron beam heating, an extended decay phase model, and a ``hot spot'' atmosphere heated by an ultrarelativistic electron beam with reasonable filling factors:  $\sim0.1$\%, 1\%, and 0.1\% of the visible stellar hemisphere, respectively.  The new modeling motivates future work to understand the origin of the extended gradual phase emission.

\end{abstract}

\section{Introduction}
In the standard flare model, coronal magnetic energy is converted to the kinetic energy of particles, which heats the chromosphere and increases the ambient charge density. This increased charge density results from direct non-thermal ionizations from the impacting flare-accelerated particles as well as thermal ionizations in the heated and compressed chromosphere from shock fronts generated by these non-thermal particles. Determining the charge density in the chromosphere from observed spectra therefore provides a way to critically test the predictions of the proposed flare energy transport and heating mechanisms, such as particle beams \citep{1971SoPh...18..489B, HF94, Abbett1999, Allred2005}, conduction \citep{Longcope2014}, and Alfv\'en waves \citep{Fletcher2008, Russell2013, Reep2016}. The charge density is a fundamental physical parameter of flare atmospheres and has many important effects on the emergent flare spectra through the recombination and bound-bound emissivity, collisional rates, and the pressure broadening of lines.  An accurate prescription for the atomic physics resulting from increased charge density in a realistic (stratified) flare atmosphere is critical for robust constraints on these heating mechanisms.  

The charge density is detectable in spectra through the symmetric broadening of hydrogen or hydrogenic-like ions caused by electric microfield pressure broadening from ambient electrons and protons\footnote{This is referred to as the linear Stark effect.  We use ``electric pressure broadening'' to refer to the linear Stark effect in hydrogen or hydrogenic ions caused by electric microfields from the surrounding distribution of electrons and protons.}.
The local electric microfield breaks the degeneracy of the orbital angular momentum states, $l$, of each principal quantum number, $n$. In both solar and stellar flares, the broadening of the hydrogen Balmer lines has often been attributed to electron pressure broadening \citep{Svestka1963, 1997ApJS..112..221J, Worden1984,HP91, Paulson2006, Allred2006, Gizis2013}, but in some flares the role of electron pressure broadening is not clear and instead directed mass flows or a large turbulent broadening is favored \citep{Doyle1988, Phillips1988, Eason1992, Fuhrmeister2011}. It is important to accurately model the electron pressure broadening to better constrain the role of other broadening mechanisms and to better understand the Balmer edge spectral region ($\lambda=3646 - 3800$ \AA), where during flares, the wings of the broadened higher order lines merge \citep{Donati1985, Kowalski2015}. 

Various theoretical frameworks have been developed for electric pressure broadening profiles \citep{Vidal1970, Vidal1973, Kepple1968}. These are convolved with the line profile (typically a Voigt) function that includes thermal and natural broadening to obtain a total line absorption coefficient \citep{Mihalas1978, Tremblay2009}. Profile convolution is computationally demanding, and analytic approximations to the true electron pressure broadening profiles were presented by \citet{Sutton1978} (hereafter, S78) who derived a broadening parameter that is added as a damping term in the Voigt profile \citep{Svestka1967}, thus significantly decreasing the computational time in radiative transfer codes. This modeling method has been shown to be adequate for the infrared lines of hydrogen in the non-flaring solar atmosphere \citep{Carlsson1992} and has been used widely in modeling flare atmospheres \citep[e.g.,][]{Kowalski2015}. The analytic approximations also have the benefit of being extended to any arbitrary high order hydrogen line, whereas the ``exact'' (theoretical) profiles are usually only available for a limited number of transitions. However, for the regime of flare chromospheric densities \citep[$n_e \gtrsim10^{13}$ cm$^{-3}$;][]{Worden1984, Donati1985, Garcia2002, Paulson2006, Kowalski2016IRIS}, there is known to be a significant discrepancy between the analytic and theoretical profiles, resulting in an order of magnitude ambiguity for the inferred charge density \citep[see][for an extensive discussion]{1997ApJS..112..221J}.  The analytic (S78) results are currently employed as Voigt profile damping parameters in several radiative-transfer codes. Notable among these are RH \citep{Uitenbroek2001}, MULTI \citep{Carlsson1986}, and the RADYN flare code \citep{Allred2015}.  In this paper, we modify the RH code to include the theoretical electron pressure broadening line profiles from the unified theory of \citet{Vidal1970, 1971JQSRT..11..263V, Vidal1973}. We show how new calculations from snapshots of dynamic flare simulations result in a significant improvement on the inferred charge density regime that is relevant for flare chromospheres.

In Section 2, we describe the electric pressure broadening theory as implemented in the RH code and compare to the analytic approximations of the line profile function. In Section 3, we show how the new broadening profiles adequately reproduce the line broadening observed in the spectrum of Vega.  In Section 4, we present improved hydrogen Balmer line broadening calculations from flare simulations with high flux electron beams that were calculated with the RADYN code, and we analyze the predictions of the new theory for the relative fluxes in each Balmer line (the Balmer decrement).  In Section 5, we compare the new broadening predictions to the spectra of the YZ CMi Megaflare and present a revised interpretation of the emission components using the spatial development of a large X-class solar flare.   In Section 6, we discuss implications for the flare heating mechanisms and avenues for future work with the new hydrogen line broadening profiles.  In Section 7, we summarize our conclusions.  In Appendix A, we present the details of two new radiative-hydrodynamic flare models that are used in the analysis.  In Appendix B, we list abbreviations and terminology.

\section{Method}\label{sec:method}
We model electric pressure broadening using the unified theory developed by \citet{1971JQSRT..11..263V, Vidal1973} (hereafter, VCS) and using the extended tables calculated by \citet{Tremblay2009}. 
Briefly, the unified theory accounts for perturbations from slowly moving (quasi-static) protons and the fast moving (dynamic) electrons with an accurate treatment of
 electron perturbations from line core to the wing.   \cite{Vidal1973} provided electric pressure broadening line profiles for the lower order transitions of hydrogen, and
\citet{Tremblay2009} (see also \citet{Lemke1997}) extended the \citet{Vidal1973} line profile function for electric pressure broadening, $S(\alpha)$, for transitions among 22 levels in the hydrogen Lyman, Balmer, Paschen and Bracket series as a function of temperature and electron density.  

We also include the non-ideal, bound-bound and bound-free opacity modifications from level dissolution following the prescription in \citet{HM88}, \citet{Dappen1987}, and \citet{Tremblay2009}.  This modeling prescription has been
included in the RH code for calculations of the Balmer edge wavelength region for flares \citep{Kowalski2015}.  In this prescription, each level of hydrogen is assigned an occupational probability, $w_{n}$ \citep[see also][]{HHL}, which 
is the probability that the level is broadened by a critical electric microfield $\beta_{\rm{crit}}$ for level $n$ and overlaps in energy with (broadened) higher energy levels.  At $\beta \ge \beta_{\rm{crit}}$, ionization\footnote{The ionization that results from level dissolution is non-degenerate pressure broadening ionization, and experiments with sodium have demonstrated that ionization occurs at  $\beta \ge \beta_{\rm{crit}}$  \citep{Pillet1983, Pillet1984, Na16}.  On the microscopic level, a Landau-Zener (L-Z) transition \citep{Zener1932} can occur between the dissolved levels $n$ and $n+1$ and rapidly proceed until the hydrogen atom is ionized \citep{Rubbmark1981, Pillet1984, HM88, Stoneman1988}; the L-Z transitions between dissolved energy levels of $n$ and $n+1$ occur at ``avoided crossings'' \citep[e.g., see the Quantum Picture described in][]{Peroti2006}.} can occur which results in the extension of the continuum opacity longward of the ionization limit.  This extended continuum opacity is often referred to as the `pseudo-continuum' opacity \citep{Dappen1987, HHL, Tremblay2009}.  As a result level dissolution, bound-free Balmer continuum flux is observed at wavelengths longer than the edge wavelength, and the higher order Balmer lines are fainter than without level dissolution. The opacity effects from level dissolution produce model spectra that are generally
consistent with the continuous flux observed at $\lambda=3646-3700$ \AA, the dissolved higher order lines from $\lambda=3700-3800$ \AA, and the blended wing flux between Balmer lines from $3700 - 3900$ \AA\ in  dMe flare spectra in the literature.  The amount by which the higher order Balmer lines dissolve (fade) into continuum flux complements the diagnostics of the
ambient flare charge density provided by the broadening of the lower order Balmer lines \citep{Kowalski2015}.

\citet{Tremblay2009} modified the 10 Balmer and Lyman calculations of VCS using $\beta_{\rm{crit}}$ as an upper limit in both the electronic broadening profile and the renormalization integral of $S(\alpha)$, because higher microfield strengths $>\beta_{\rm{crit}}$  have a significant probability to transform bound-bound opacity into bound-free opacity \citep{Seaton1990}. This makes the line broadening theory fully consistent with the independent \citet{HM88} equation-of-state.  We have calculated the Balmer and Lyman lines up to
H20, and for 19 Paschen lines and 10 Brackett lines using $\beta_{\rm{crit}}$ \citep[cf Equation 18 of][]{Tremblay2009}. Hereafter, we refer to these profiles calculated with the VCS unified theory extended with the modifications of \citet{Tremblay2009}, the occupational probability formalism of \citet{HM88}, and the non-ideal (pseudo-continuum) opacity of \citet{Dappen1987} as the TB09$+$HM88 profiles.  

\begin{figure}[htbp]
\plotone{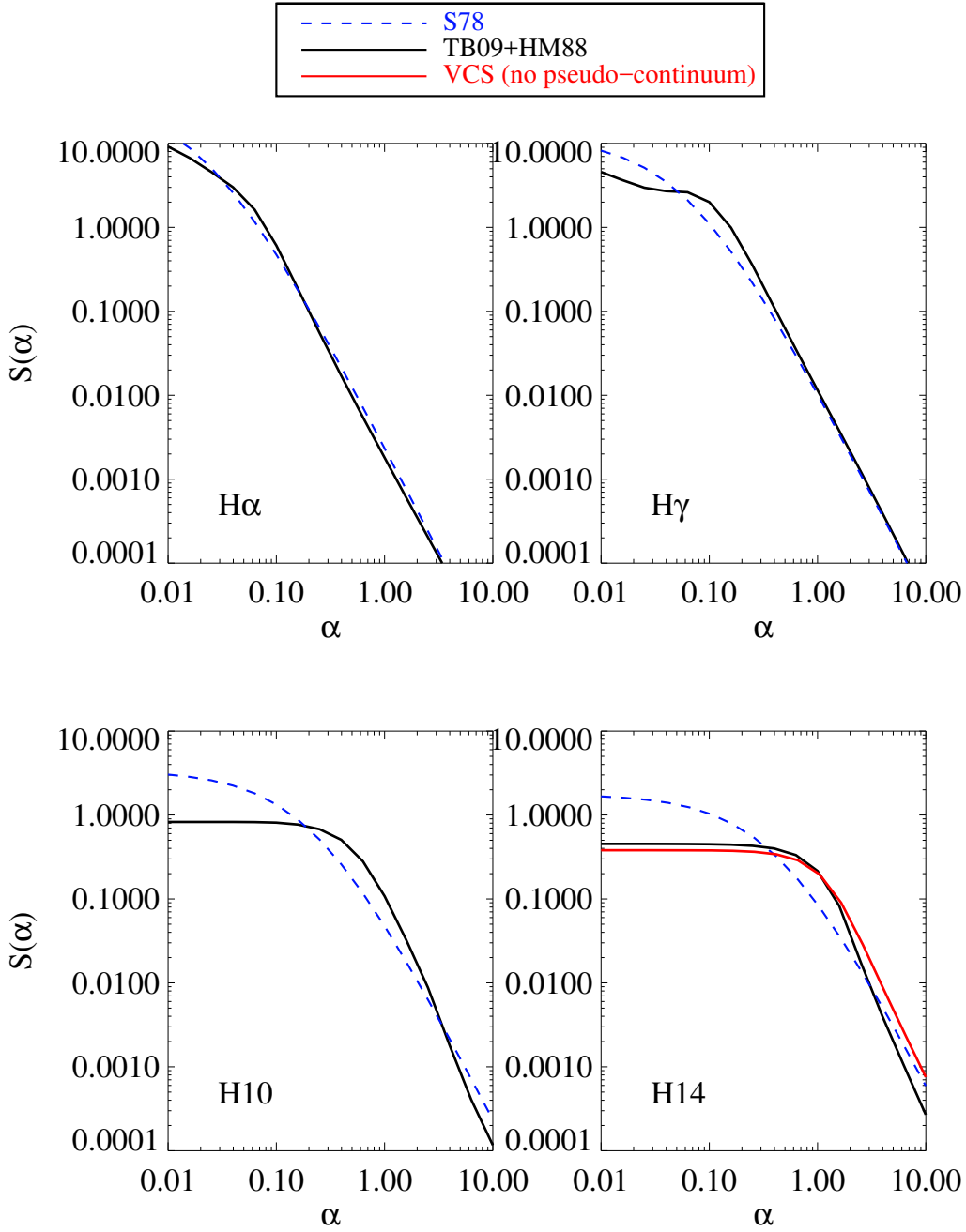}
\caption{ Electric pressure broadening profiles S($\alpha$) comparing the S78 method (dashed blue) and TB09$+$HM88 method (solid black) for select Balmer lines.  $\alpha$ is proportional to $\lambda-\lambda_o$; the conversion from $\alpha$ to $\Delta \lambda$ in \AA\ is $1.25\times10^{-9} \times n_e^{2/3}=[0.58, 2.69, 12.5, 37]$ for $n_e=[10^{13}, 10^{14}, 10^{15}, 5\times10^{15}]$ cm$^{-3}$.   These profiles use a temperature of $T=10,000$ K and $n_e=10^{14}$ cm$^{-3}$. The H14 profile (solid red) is the VCS profile of H14 from \citet{Lemke1997} that does not account for level dissolution ($\beta_{\rm{crit}}$ is not used in the line profile calculation; see text here and \citet{Tremblay2009}). }   \label{fig:vcsprofiles}

\end{figure}

 In Figure~\ref{fig:vcsprofiles} we plot several of the TB09$+$HM88 Balmer line profiles and compare them with the corresponding line profiles from S78 (Method \#1), where $\alpha = (\lambda-\lambda_o)/F_o$ ($\lambda$ is expressed in \AA) and $F_o$ is the field strength at the average interparticle separation.  The conversion from $\alpha$ to $\Delta \lambda$ in \AA\ is $F_o=1.25\times10^{-9} \times n_e^{2/3}=[0.58, 2.69, 12.5, 37]$ for $n_e=[10^{13}, 10^{14}, 10^{15}, 5\times10^{15}]$ cm$^{-3}$.  The VCS (no pseudo-continuum) profiles S($\alpha$) for different values of $n_e$ generally overlap, and the comparisons to the S78 profiles were discussed in S78 for the lower order Balmer lines. Here we show the comparisons to the TB09$+$HM88 profiles only for $n_e=10^{14}$ cm$^{-3}$, which demonstrates that the entire profile for H$\alpha$ is 
well-reproduced by the S78 approximations and that the asymptotic Holtsmark profile ($\propto \alpha^{-5/2}$) is well-reproduced for the H$\gamma$ line.  However, 
for the higher order Balmer line profiles (H10, H14) in Figure \ref{fig:vcsprofiles} the discrepancy becomes large \citep[see also Figure 7 of][for a comparison of other electric pressure broadening theories to S78 for H10]{1997ApJS..112..221J}.  We show the H14 profile calculated with $\beta_{\rm{crit}}$ compared to the profile from \citet{Lemke1997} without using $\beta_{\rm{crit}}$ (VCS, no pseudo-continuum).  For larger electron densities, the discrepancy between the VCS (no pseudo-continuum) H14 profile and the TB09$+$HM88 H14 profile is even larger, as expected \citep{Tremblay2009}.

\begin{figure}[htbp]
\plotone{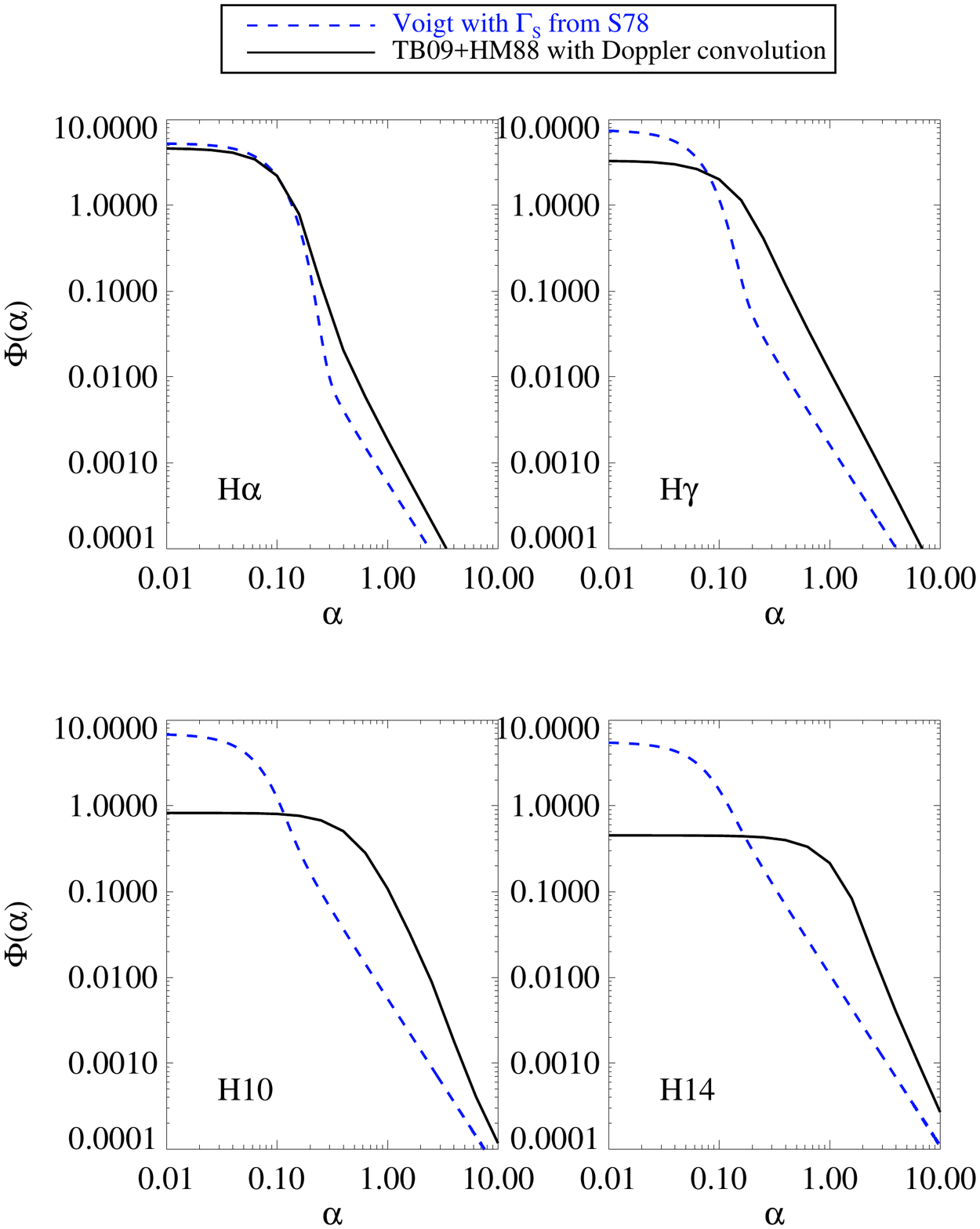}
\caption{ Line profiles $\Phi$($\alpha$) comparing the Voigt function with $\Gamma_{\rm{S}}$ obtained from S78 (dashed blue) to the TB09$+$HM88 profiles with Doppler convolution (solid black) for select Balmer lines.   $\alpha$ is proportional to $\lambda-\lambda_o$; the conversion from $\alpha$ to $\Delta \lambda$ in \AA\ is $1.25\times10^{-9} \times n_e^{2/3}=[0.58, 2.69, 12.5, 37]$ for $n_e=[10^{13}, 10^{14}, 10^{15}, 5\times10^{15}]$ cm$^{-3}$. These profiles use a temperature of $T=10,000$ K and $n_e=10^{14}$ cm$^{-3}$.  The TB09$+$HM88 profiles have been calculated with $\beta_{\rm{crit}}$ as described in \citet{Tremblay2009}.  The line profile functions largely differ for the two methods in the far wings of all Balmer lines and in the wings and cores for the Balmer lines that are higher order than H$\alpha$. }   \label{fig:vcsprofiles2}
\end{figure}

The analytic approximations from S78 are often used to estimate an electric pressure damping parameter, $\Gamma_{\rm{S}}$, that is added to the total damping parameter in the Voigt profile ($H$), which 
dramatically decreases computational time and increases flexibility in radiative transfer codes since an S78 profile can be calculated for a transition with any arbitrarily large upper principal quantum level.  In Figure \ref{fig:vcsprofiles2} we plot the normalized line profile function $\Phi(\alpha)$ for several Balmer lines 
using a Voigt function with $\Gamma_{\rm{S}}$ from S78 in the Voigt profile ($\Phi(\alpha)\propto H(\Gamma_{\rm{tot}}=\Gamma_{\rm{tot}}+\Gamma_{\rm{S}})$)  compared to the the TB09$+$HM88 line profile functions with a Doppler convolution ($\Phi(\alpha) \propto S(\alpha)*H$).
We use $T=10,000$ K for Doppler broadening in this comparison.  Here it is most striking that using $\Gamma_{\rm{S}}$ in the Voigt profile does not adequately produce the broadening in the Balmer lines 
as predicted by the TB09$+$HM88 profiles.  

We modified the RH radiative transfer code to convolve the Voigt function (with thermal and natural broadening) with the TB09$+$HM88 profiles ($S(\alpha)$), following 
\citet{Mihalas1978} and \citet{Tremblay2009} to obtain the normalized line profile function:

\begin{equation} \label{eq:convolution}
\Phi(\alpha) = \frac{d\nu}{d\alpha} \integ{-\infty}{\infty}{S^*(\Delta \nu+ v \Delta \nu_D)H(a,v)}{v}
\end{equation}

\noindent where $S^*$ is $S(\alpha)$ converted to frequency units and normalized to 1 from $-\infty$ to $+\infty$.

For the model atmospheres calculated in this work, we use a 20 level hydrogen model atom with
the TB09$+$HM88 profiles for the Balmer and Paschen lines as well as the opacity effects from level dissolution at the Balmer and Paschen edges. 
We also use a 5-level Ca II model ion $+$ continuum in the non-LTE calculation\footnote{We use the Ca II model that comes with the standard RH distribution.  The only modification that is necessary is to extend the wavelength grid to cover the far wings of Ca II H and K.}.  The Ca II H and K lines are important to include for a detailed assessment of the blending of the hydrogen line wings from $\lambda=3900-4000$ \AA.


\section{Comparisons to Vega}\label{sec:vega}
\begin{figure}[htbp]
\plotone{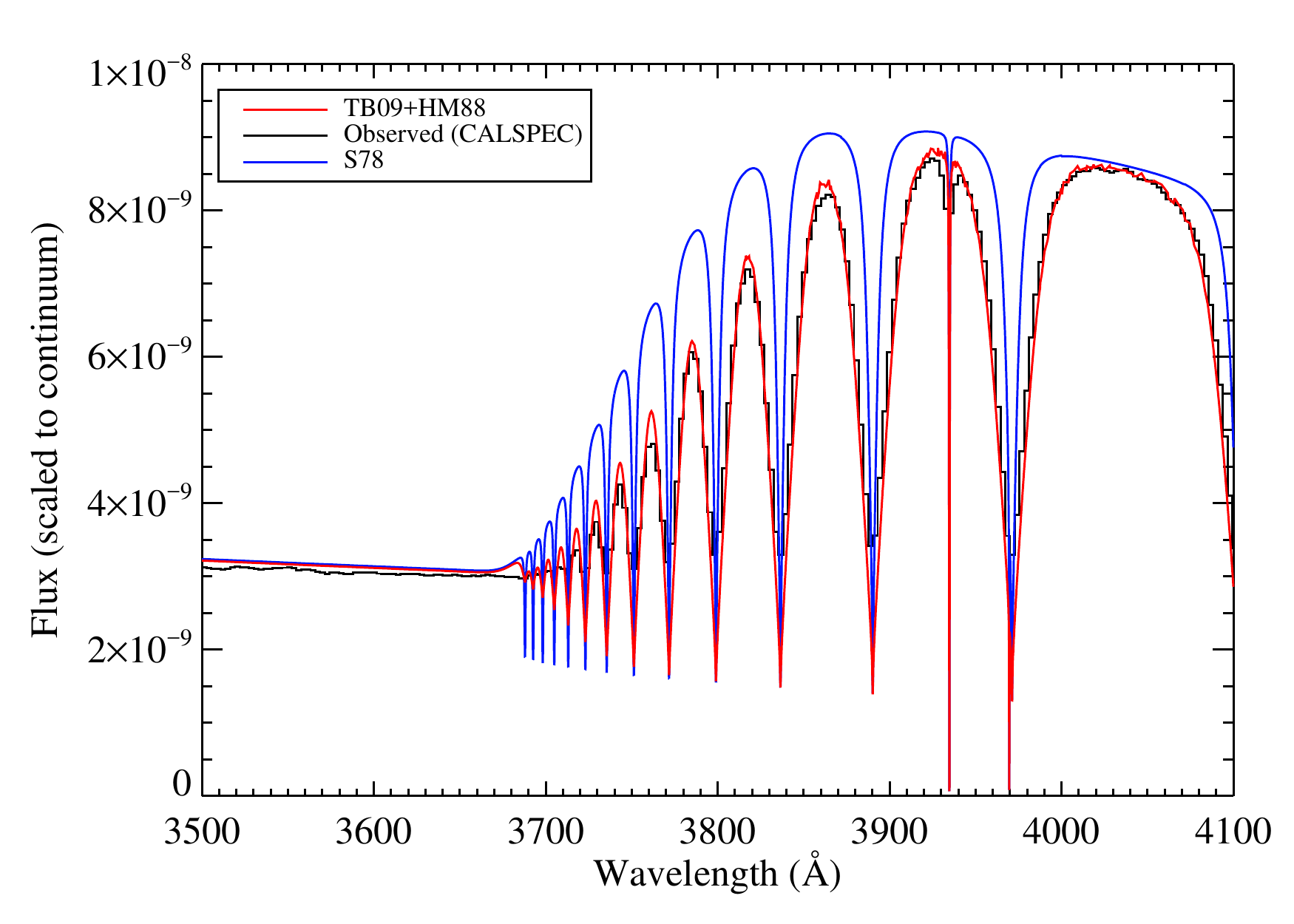}
\plotone{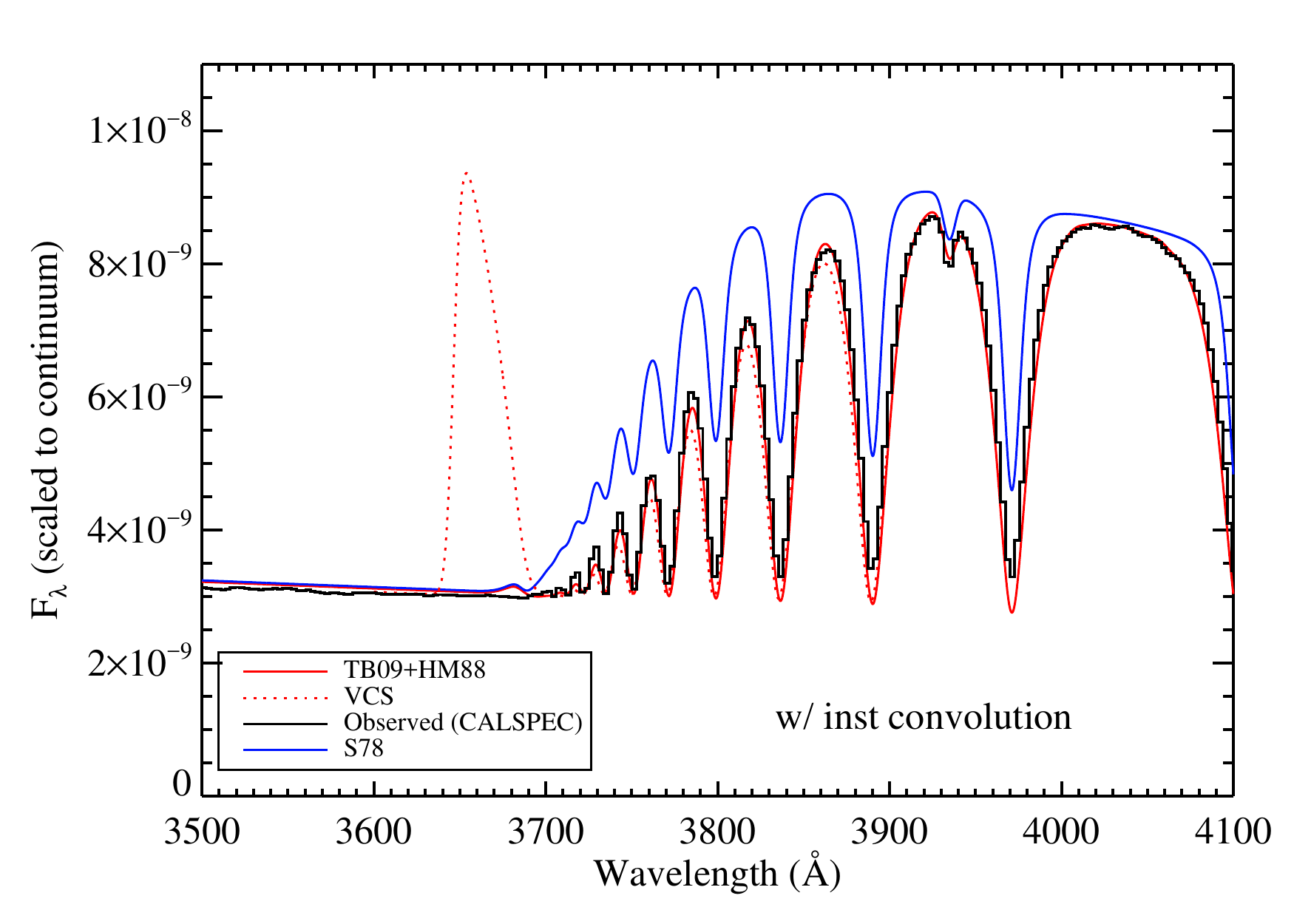}
\caption{(Top) Observed spectrum of the A0 star, Vega, (black line) compared to predicted spectra using the S78 method (blue line) and the TB09$+$HM88 method (red line) from $\lambda=3500-4100$ \AA. (Bottom) The model spectra in the top panel have been convolved by the instrumental resolution of FWHM $=$ 8 \AA.  The TB09$+$HM88 prescription for broadening and level dissolution (solid red) better accounts for the broadening and depth of the hydrogen Balmer lines.  The dotted line shows the prediction using the VCS profiles without the bound-bound and bound-free opacity effects from level dissolution (no pseudo-continuum). \label{fig:vega}}
\end{figure}

To test our method, we compare observations of Balmer line flux from the A0 star, Vega, with simulated spectra produced from RH using our method and that of S78. The observations of Vega are described in \citet{Bohlin2004} and \citet{Bohlin2007} and were obtained from the Space Telescope Science Institute CALSPEC Calibration database.   We have chosen Vega as a test case because it is a well-studied star with a known atmospheric structure and electron densities in the region of Balmer line formation that are similar to those previously inferred in flaring chromospheres.
 We input the atmospheric structure of Vega obtained from the ATLAS9 grid \citep{Castelli1994, Castelli2004} into RH.   
 Figure~\ref{fig:vega} shows the results of this comparison (where the bottom panel shows the effect of instrumental convolution). The black line is the observed spectrum, and the red and blue lines are the simulated spectra using our method with the TB09$+$HM88 profiles and the method of S78, respectively. Clearly, the TB09$+$HM88 prescription for broadening and level dissolution produces a much better fit to the observed spectrum at $\lambda \gtrsim 3700$ \AA. 

The TB09$+$HM88 broadening has been shown to accurately reproduce the CALSPEC spectra
of Vega and white dwarfs using other radiative transfer codes \citep{Bohlin2014}.
Vega is a rapidly rotating star that is seen nearly pole-on \citep{VegaAuf}.  We do not include the projected rotational velocity,
which broadens only the line core region.  We investigate the effect on the wing broadening from the pole-to-limb variation of effective temperature ($\Delta T_{\rm{eff}} \sim 2200$ K) using the ATLAS9 grid.  A model with  $T_{\rm{eff}} = 10250$ K at $\mu=0.95$ (representative of the pole) together with a model with $T_{\rm{eff}} = 8000$ K at $\mu=0.05$ (representative of the limb)
does not improve the broadening produced by the S78 prescription.  

As described in Section
\ref{sec:method}, we included the 
prescription from \citet{Dappen1987} for bound-bound and
bound-free opacity effects due to hydrogen level dissolution into RH with the 
TB09$+$HM88 profiles\footnote{In \cite{Kowalski2015}, the same prescription with the S78 electric pressure broadening 
for flare model atmosphere calculations with the RH and RADYN codes was used.}. In Figure \ref{fig:vega} (top), the 
Balmer lines from S78 are narrower than with the VCS theory as implemented in the TB09$+$HM88 profiles, and the
 blending of the wings of the
Balmer lines between $\lambda=3700-3950$ \AA\ is not nearly as prominent as the calculations with the TB09$+$HM88 profiles, which result in deeper absorption 
and more blending in the wings even when instrumental resolution is incorporated (bottom panel of Figure \ref{fig:vega}).  
Both prescriptions produce bona-fide Balmer continuum flux longward of the Balmer limit at $\lambda=3646$ \AA.

In the bottom panel of Figure \ref{fig:vega}, the red dotted line 
shows the calculation without the opacity effects from level dissolution.  Specifically, we set the occupational probability of each
level $n$ for hydrogen equal to one, we set the Balmer opacity longward of the Balmer limit to zero (no pseudo-continuum), and we use the VCS line broadening profiles from \citet{Lemke1997} 
which do not include $\beta_{\rm{crit}}$ in the calculation of the line profile shape.   We refer to this calculation as the ``VCS (no pseudo-continuum)''.   The
difference between this calculation compared to the observation at $\lambda \sim 3650 - 3690$ \AA\ demonstrates that the bound-free opacity from level dissolution (the pseudo-continuum opacity) is necessary to accurately account for the continuum flux at these wavelengths even with the deep absorption and large broadening produced by the TB09$+$HM88 profiles.  For the charge density that produces the broadening in Vega, the Balmer bound-free opacity at wavelengths longward of the Balmer limit generally follows the trend in the flux between the Balmer lines in the S78 emergent flux spectrum at $\lambda = 3700 - 3850$ \AA\ in Figure \ref{fig:vega} (top).  At these wavelengths in the TB09$+$HM88 prediction, the dominant opacity is the superposition of the bound-bound opacities of the hydrogen lines\footnote{The TB09$+$HM88 profiles are calculated using $\beta_{\rm{crit}}$ because of the level dissolution from ambient protons;  the profiles extend to infinity because
the level dissolution from electrons does not have a sharp cutoff \citep[see the discussion in HM88 and][]{Tremblay2009}.}, which causes the 
calculation with the TB09$+$HM88 profiles and the calculation with the VCS (no pseudo-continuum) profiles to be similar at these wavelengths but with broader higher order line profiles in the latter \citep{Tremblay2009}.  
 The level dissolution opacity (i.e., that which is often referred to as the pseudo-continuum opacity) is most critical for reproducing the flux of Vega at $\lambda$ between 3646 \AA\ and 3700 \AA, where the oscillator strength density is very low and the Balmer lines are completely dissolved \citep{Dappen1987}.  

\begin{figure}
\gridline{\fig{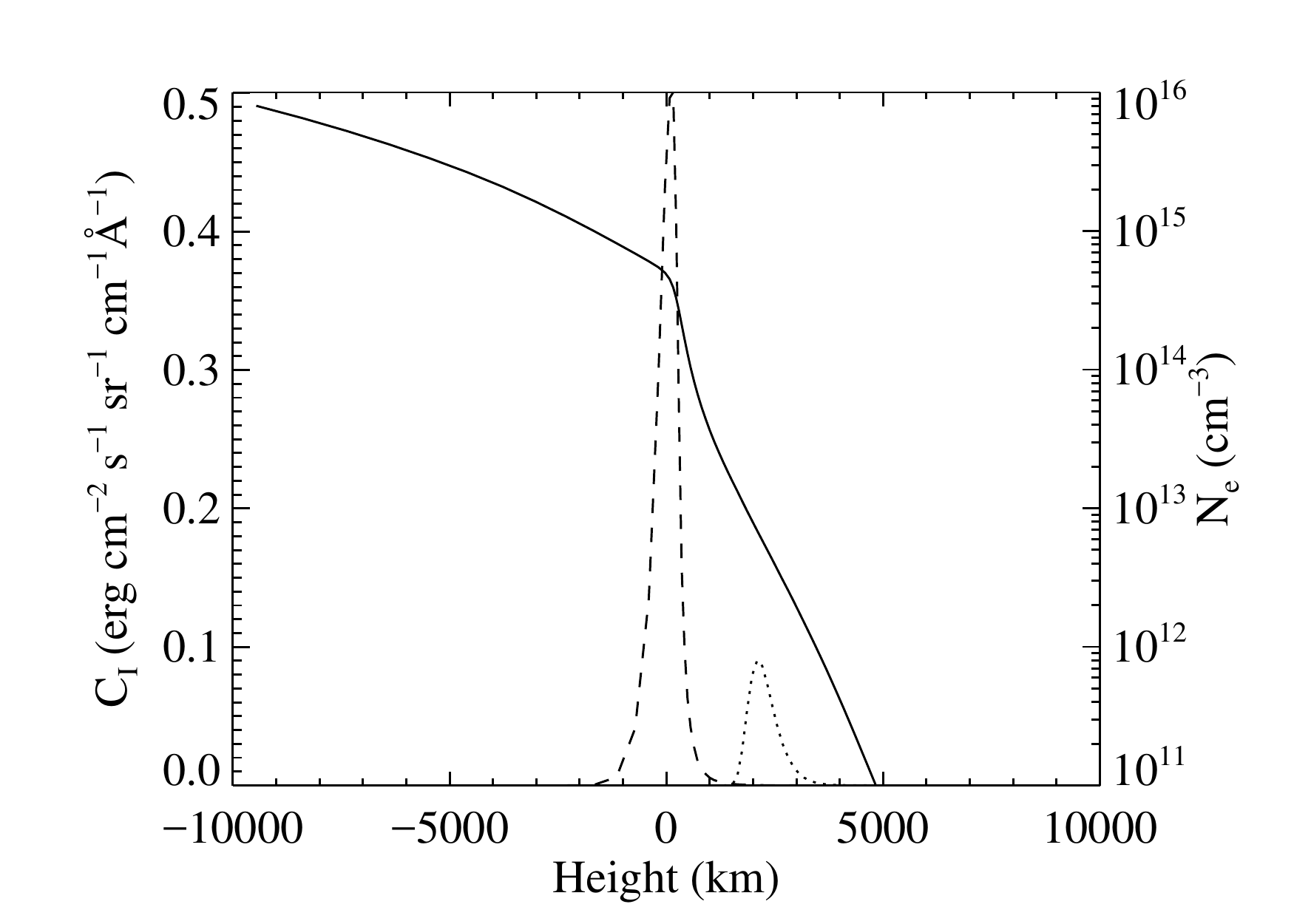}{0.47\textwidth}{}}
\gridline{\fig{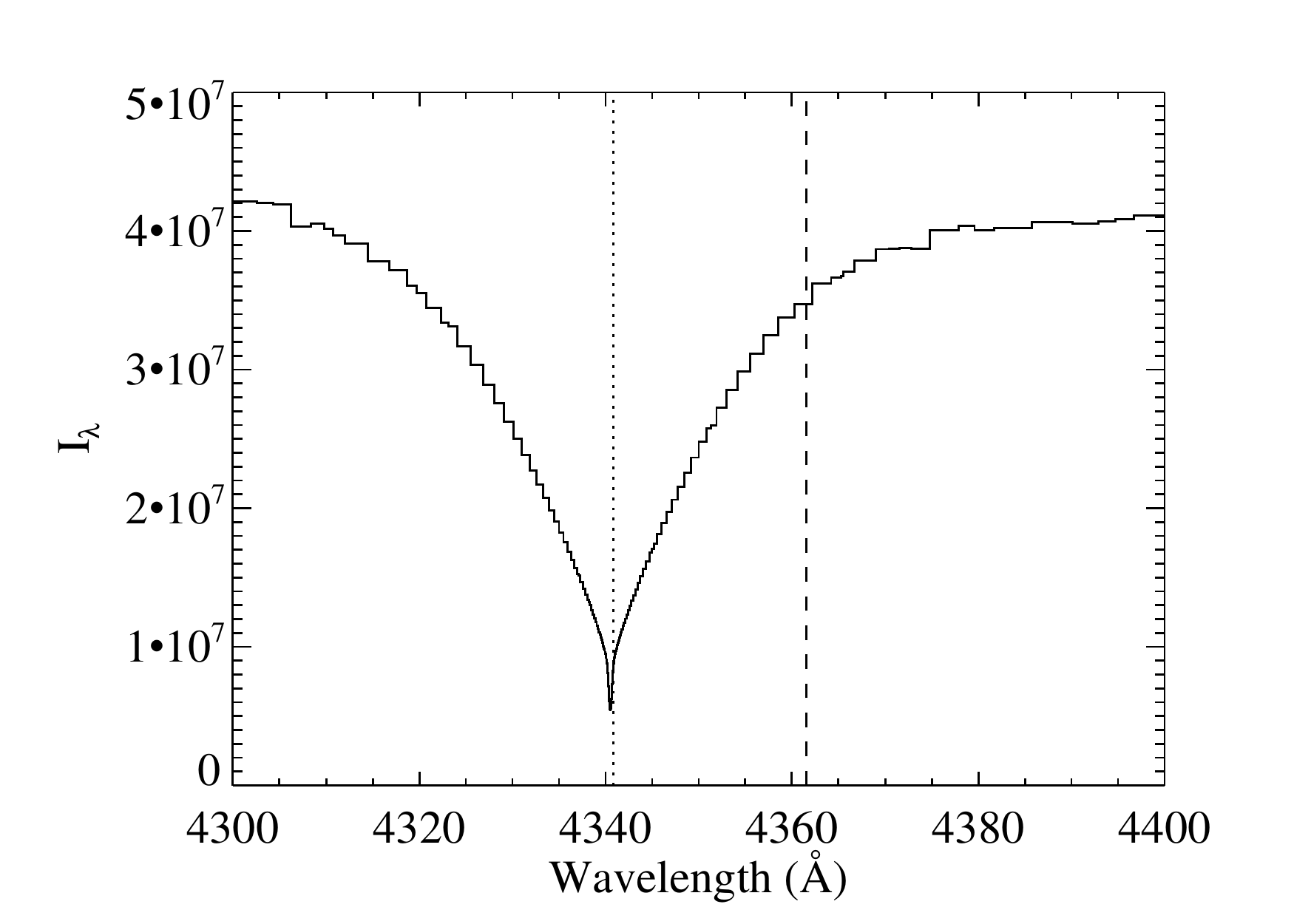}{0.47\textwidth}{}}
\caption{(Top panel) Contribution function plotted as a function of height above the photosphere for two representative wavelengths in the H$\gamma$ line. These are at line center (dotted line) and in the far wing at $\lambda_{\rm{rest}}+20$ \AA\ (dashed line). The electron density (solid line) measured from the right axis is also plotted. (Bottom panel) The H$\gamma$ line profile with dotted and dashed lines indicating the wavelengths where the corresponding contribution functions are calculated. The line forms over a large range of electron density and the emergent spectrum is a combined intensity profile over a large range of electric pressure broadening amounts.\label{fig:vegacf}}
\end{figure}

As an example of using the predicted spectra to estimate the electron density in the region of Balmer line formation, we plot in Figure~\ref{fig:vegacf} the line profile and contribution function for the H$\gamma$ line calculated with the TB09$+$HM88 profiles in RH. The contribution function indicates where emitted line intensity originates as a function of atmospheric height and wavelength, such that integrating the contribution function over height reproduces the line profile. In the top panel of Figure~\ref{fig:vegacf} we plot the contribution function for H$\gamma$ at line center (dotted line) and in the far wing (dashed line). The wavelengths where these contribution functions are measured are indicated with corresponding vertical lines in the line profile plot in the bottom panel. The solid line in the top panel shows the electron density measured on the right axis. Together these indicate that the intensity near line center forms in a region with electron density ranges from $\sim \sn{3-7}{12}$ cm$^{-3}$ and the far wings ($\lambda_c +20$ \AA) form in a region with $\sim \sn{4-5}{14}$ cm$^{-3}$.

\section{Balmer Line Broadening in Stellar Flares\label{sec:widths}}
In our previous work \citep{Allred2005, Allred2006, Kowalski2015,
  Kowalski2016, Kowalski2016IRIS} we have used the RADYN code
\citep{Carlsson1997} to simulate the dynamics of flaring loops in 
solar and dMe atmospheres. We refer the reader to \citet{Allred2015}
for an extensive description of the RADYN flare code.

The widths of the Balmer lines are often used
to compare to models of pressure broadening \citep{Svestka1963,
  Svestka1967} to infer an electron density, while the profile shapes in the
line wings have been compared to theoretical model predictions \citep{Doyle1988, Eason1992,
  Allred2006, Paulson2006}.  The highest order Balmer lines that are
resolved have been used with the Inglis-Teller relation to probe the
charge density in flares \citep{Kurochka1970, HP91}.
  To account for self-consistent optical depth,
density, and temperature
variations over the regions of line formation, we use our modeling method to
revisit the broadening predictions from recent RHD flare models of
dMe flares that produce large values of the electron density
($n_e\gtrsim10^{15}$cm$^{-3}$).
These comparisons can be extended to solar flare model atmospheres
that exhibit lower electron
densities values of $n_e\sim10^{13}-5\times10^{14}$ cm$^{-3}$
\citep{Donati1985, 1997ApJS..112..221J, Kowalski2016IRIS} to make
predictions for the Daniel K. Inouye Solar Telescope (DKIST).

\subsection{Instantaneous Model Predictions} \label{sec:stat}
In \cite{Kowalski2015} and \cite{Kowalski2016} we presented model
flare atmospheres heated by a high flux of nonthermal electrons
($10^{13}$ erg cm$^{-2}$ s$^{-1}$; F13) which produce a 
$T\sim10^4$ K blackbody-like continuum
flux distribution as observed in many flares from active M dwarf
stars.  The hot blackbody-like continuum distribution results from 
large continuum optical depths from a dense, heated chromospheric
condensation.  The charge density in the continuum-emitting layers
achieves a maximum value of $\sim 5\times10^{15}$ cm$^{-3}$.  The
 opacity effects from level dissolution produces Balmer
continuum flux\footnote{The Balmer continuum flux from level dissolution is
    sometimes also referred to as the pseudo-continuum
  flux, or as the L-Z continuum flux in \citet{Kowalski2015}.} at wavelengths longer than the Balmer limit (3646 \AA)
and higher order Balmer emission lines that fade into this continuum flux.
These properties have been observed in high resolution spectra of a
large dMe flare \citep{Fuhrmeister2008}.  With the S78 approximations as Voigt profiles in the RH
code, the highest order Balmer lines that are predicted at
this charge density are H10/H11, which is reasonable compared to 
the observations of some dMe flares \citep{Kowalski2013,Kowalski2016}.  

\begin{figure}
\plotone{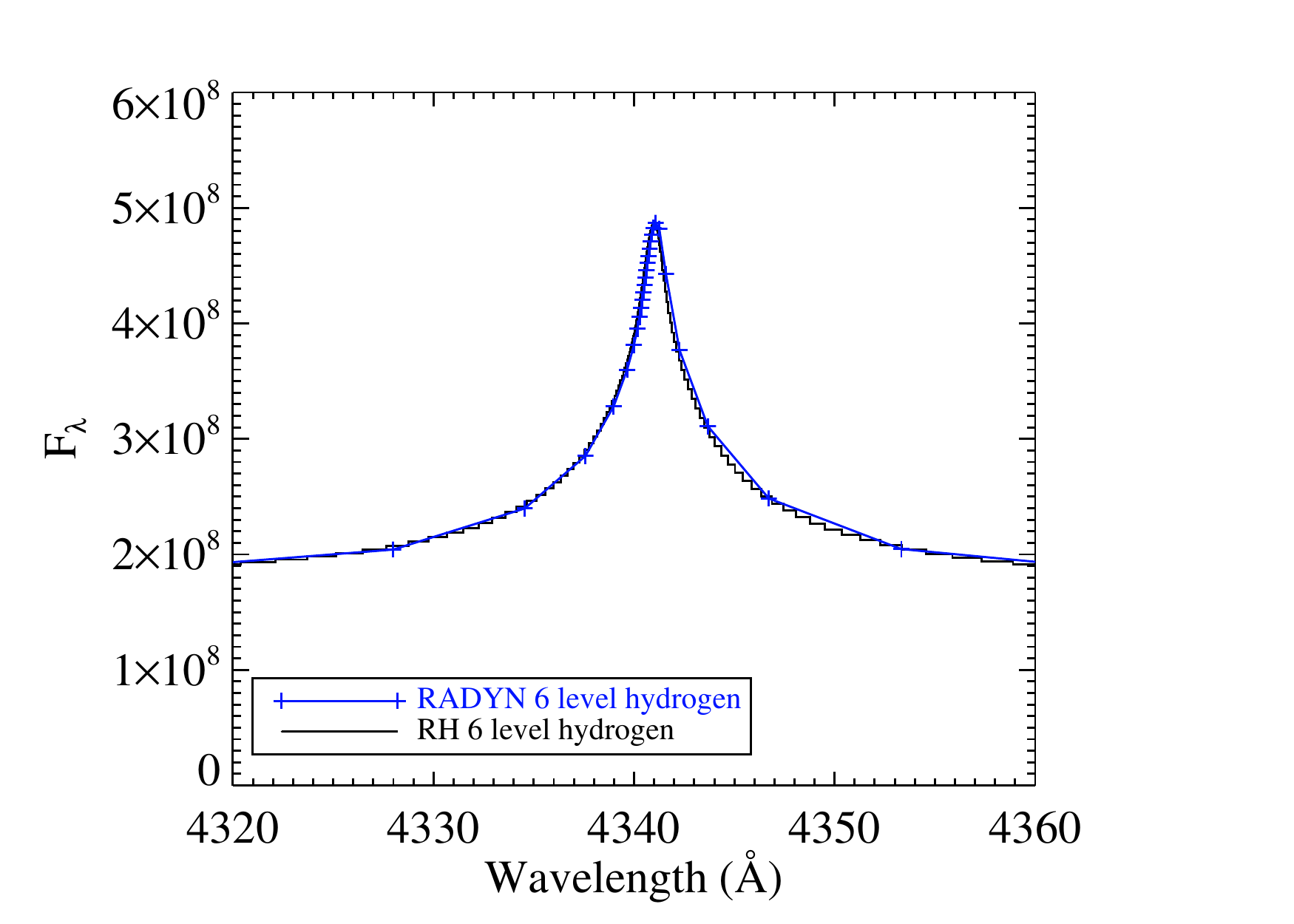}
\caption{ Comparison of the F13 dpl at $t=2.2$~s H$\gamma$
  line profile calculated with RADYN (with non-equilibrium ionization and
  excitation) and RH (with non-equilibrium ionization).  The statistical
  equilibrium assumption for excited levels of hydrogen employed in RH
  does not have an effect on the
  line profile (at this time step).  In this comparison, both calculations use the S78
  prescription for electric pressure broadening. 
\label{fig:rhradyn}}
\end{figure}

Because of the computation time required for the convolution (Equation
\ref{eq:convolution}), it is unfeasible to incorporate the TB09$+$HM88
method directly into
RADYN. Instead, we input snapshots of the atmospheric state from RADYN
into RH. It is well-known that the flaring atmosphere is very
dynamic.  The dynamics and the proper time-dependent ionization
  (electron density) at each snapshot are
included, but
statistical equilibrium for the excitation is assumed in solving the
radiative transfer equation in the RH calculation.  To
understand the extent of this limitation, we plot Balmer lines modeled
using RADYN and RH in Figure \ref{fig:rhradyn}.  For this comparison, RH has been configured
to most closely match the radiative transfer method in RADYN, (i.e.,
using a 6-level hydrogen atom, complete redistribution, and S78 
broadening).  In this configuration, the only difference between the
radiative transfer methods employed by these codes is the assumption
of statistical equilibrium in the excited levels employed in RH. Figure \ref{fig:rhradyn}
indicates that, for the $t=2.2$~s snapshot in the F13 model from \citet{Kowalski2015}, the statistical equilibrium
limitation is a small effect, justifying our use of RH.  The same is
true for $t=2.2$~s of the F12 model presented in \citet{Kowalski2015}.

In RH, the occupational probability formalism is not included in
  the rate equations for statistical equilibrium
  as described in detail in \citet{HHL}.  In our implementation,
  we include the occupational probability prescription only in the
  NLTE bound-bound and bound-free opacity and emissivity
  \citep[Section \ref{sec:method}, see also][]{Kowalski2015}.  
  Including occupational probabilities in the rate equations for
  statistical equilibrium in RH would likely affect the electron
  density and the populations of the upper levels of hydrogen.  At
  this expense, we use the
  non-equilibrium electron density from a snapshot from RADYN.  
 The agreement with the observed spectrum of Vega (Section
 \ref{sec:vega}) justifies that the method currently employed in RH is
 sufficiently accurate for atmospheres that are near LTE conditions in
 their continuum-emitting layers, as is the case for the F13 flare models at
 $t=2.2$~s \citep[see the discussion in][]{Kowalski2015}.

\begin{figure}
\plotone{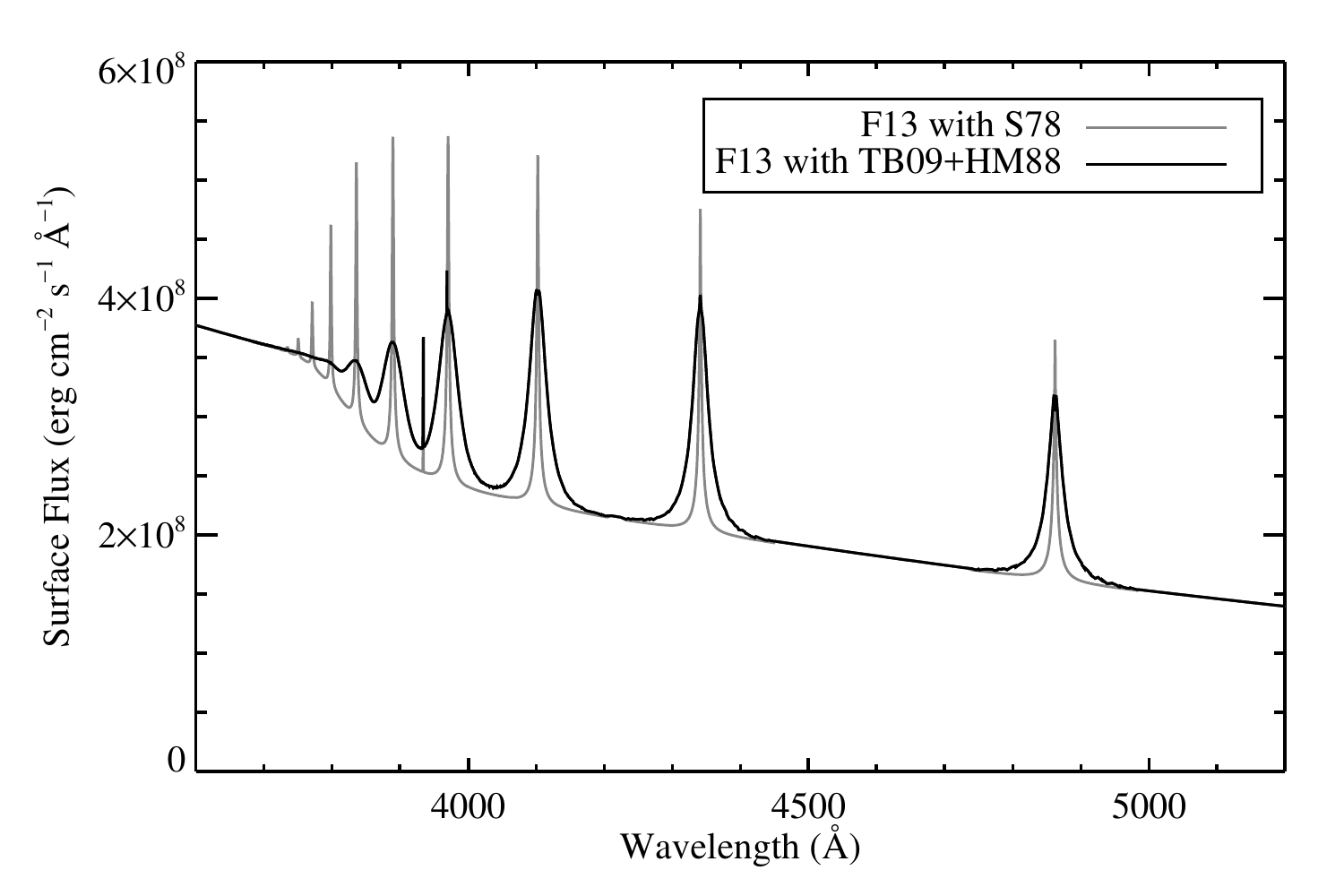}
\caption{ A flare model spectrum (F13 $\delta=3$ at $t=2.2$~s) from
  $\lambda=3600-5200$ \AA\ calculated with RH with the TB09$+$HM88
  broadening compared to the S78 broadening.  The S78 spectrum is obtained
  from the F13 model in \citet{Kowalski2016} using the method of \citet{Kowalski2015}.
The opacity effects from dissolved
  levels are included in both calculations.  
  Instrumental convolution has not been applied here.  The TB09$+$HM88
  profiles produce much broader Balmer lines than S78.  
  In the TB09$+$HM88 calculation, the higher order Balmer lines fade into
  the dissolved level continuum
  flux at a longer wavelength than the S78 prescription.  
\label{fig:f13}}
\end{figure}

We repeat the RH calculation with the TB09$+$HM88 profiles convolved
with the Voigt function
for the $\delta = 3$, F13 model atmosphere at $t=2.2$~s
from \cite{Kowalski2016} and for the double power-law ($\delta=3$ at
$E < 105$ keV, $\delta=4$ at $E>105$ keV) F13 model at $t=2.2$~s (hereafter, F13 ``dpl'') from
\citet{Kowalski2015}.  The new 
prediction in the Balmer jump wavelength region is shown in Figure
\ref{fig:f13} for the F13 $\delta=3$ at 2.2~s, which exhibits broader Balmer lines while
H9 is the highest
order Balmer line that is not completely transformed into Balmer
continuum flux longward of the Balmer edge.  With the S78 method,
H11 is the bluest detectable (undissolved) Balmer line in
emission. We use the TB09$+$HM88 profiles which have been calculated with
$\beta_{\rm{crit}}$ but find that these increase the peak of H8 Balmer
line and decrease the trough between H8 and H9 by $\sim2$\% compared
to a calculation without $\beta_{\rm{crit}}$ (VCS, no psuedo-continuum).
Using the TB09$+$HM88 profiles, the H10 and H11 lines are broadened much more
than with the S78 prescription (Figure \ref{fig:vcsprofiles}); the occupational probability for the 
upper levels of these transitions are small which  
 causes the flux at these wavelengths to be 
dominated by dissolved level continuum flux in the emergent spectrum.
Higher order Balmer lines in the impulsive phase that are as faint
(relative to the nearby continuum) as the new F13 prediction are very
rare \citep[e.g.,][]{Garcia2002}.

We convolve the F13 model spectra in Figure \ref{fig:f13} with a representative spectral resolution ($R\sim450$) from the flare spectral atlas of
\cite{Kowalski2013} and calculate the width at 10\% of maximum line
flux (0.1 width) for the H$\gamma$ line to be 78 \AA\
for the TB09$+$HM88 prescription and 32 \AA\ for the S78 prescription.  For
the F13 dpl at $t=2.2$~s, the 0.1 width of H$\gamma$ is larger,
$\sim$100 \AA\ with the TB09$+$HM88 broadening.
Typical line widths in dMe flares are 15-20 \AA\ for large flares at
moderate spectral resolution \citep{HP91}; the
maximum
observed 0.1 widths for the H$\gamma$ line at low spectral resolution are 
$30-40$ \AA\ \citep[Figure 4.13 of][]{Kowalski2012} at times of
brightest line emission, but
the line broadening has been observed to be larger ($45-50$ \AA) in the
mid-rise phase of large flares \citep[Figure 4.14
of][]{Kowalski2012}.  The contribution function-weighted charge density
over which the H$\gamma$ line forms is $\sim 3.5-5 \times10^{15}$
cm$^{-3}$ for the F13 $\delta=$3 (and $4-6 \times10^{15}$
cm$^{-3}$ for the F13 dpl) at $t=2.2$~s.
The F13 model with TB09$+$HM88 profiles demonstrates that the F13
instantaneous spectrum at 2.2~s, and therefore this range of charge density, is not consistent with even the
largest observed values of the broadening in flares.  In Section
\ref{sec:multithread}, we show that adding flux spectra with lower
broadening (at earlier
and later times in the heating simulation) to emulate simultaneously heating and
cooling loops produces a broadening that
is closer to the observations.  
 Because of the larger broadening with the TB09$+$HM88 profiles, the emergent
 line intensity in the F13 model originates from a larger physical depth range in the
 atmosphere, and the wavelength-integrated emergent
line flux of the Balmer lines is therefore also several times ($\sim$2.5x)
larger than in the case of the S78 profiles.

\begin{figure}
\plotone{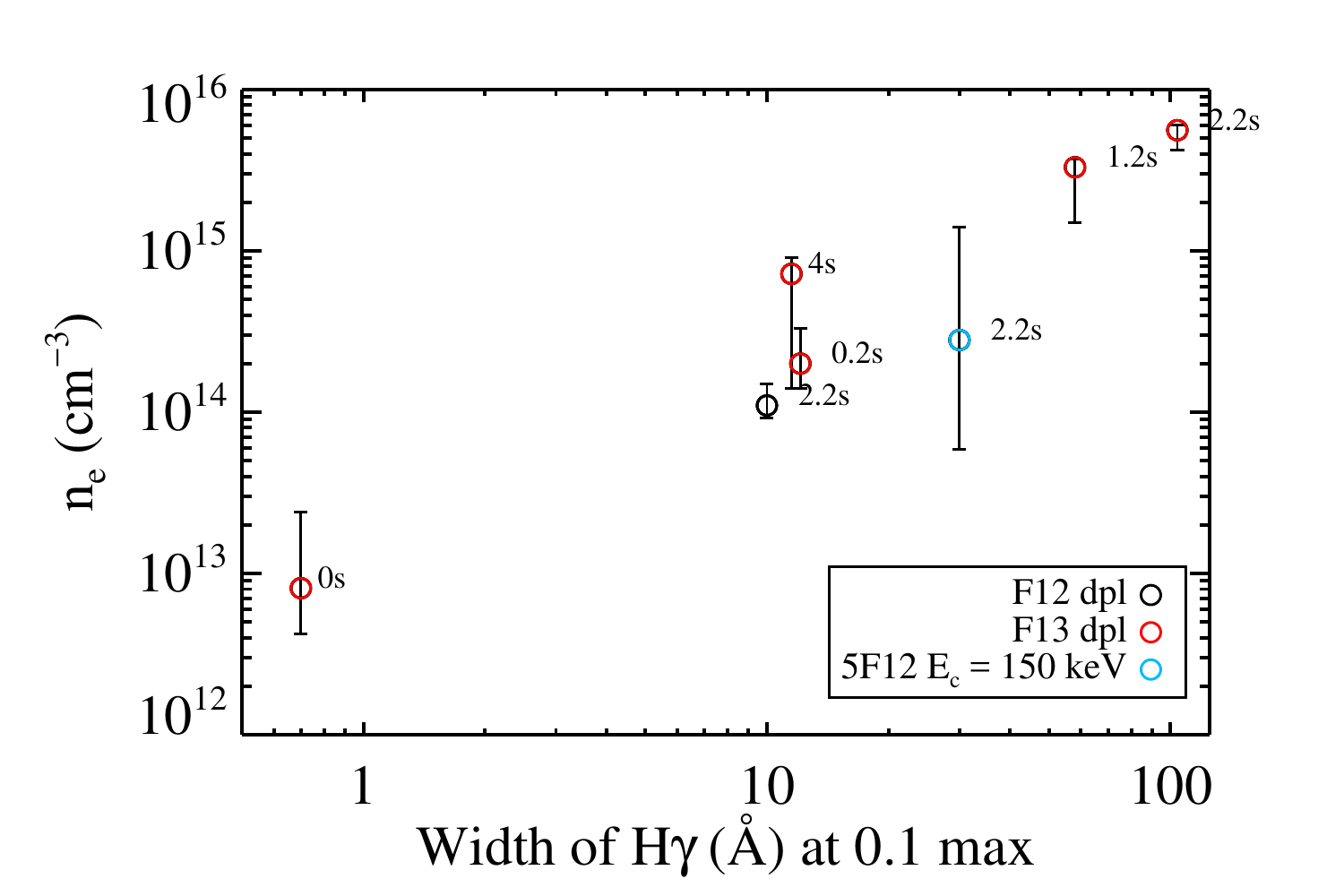}
\caption{ The electron density weighted by the contribution function
  to the emergent intensity ($\mu=0.95$) vs. the 0.1 width of the
  H$\gamma$ line (without instrumental convolution).  
  The values are shown for representative times during the F13 dpl flare
  simulation, for the F12 dpl simulation (at $t=2.2$~s), for our
  pre-flare M dwarf model atmosphere, and for a new flare simulation
  with an energy flux of 5F12, $E_c=150$ keV, and $\delta=3$.  The circles show the mean
  electron density over the line, and the upper and lower error bars show the
  maximum and minimum electron density over the line.  The ``far
  wing'' values described in the text refer to the value of the
  contribution function-weighted electron density at $\lambda_{\rm{rest}}+20$
  \AA, which approximately corresponds to the upper error bar.  There is a
  general trend with slope $\approx$2/3 for the snapshots from the F13 simulation.  
\label{fig:nel}}
\end{figure}

In Figure \ref{fig:nel}, we plot the widths of the H$\gamma$ line
for the evolution of the F13 dpl simulation compared to 
the width of the line from the F12 dpl simulation from \citet{Kowalski2015} in order to bracket the range
of
electron densities relevant for flares
($10^{13}-5\times10^{15}$ cm$^{-3}$).   For the F13 dpl model evolution in
Figure \ref{fig:nel}, the slope of the relation is
approximately 2/3 which is the expected scaling of the wavelength
shift from electric pressure broadening \citep[see e.g.,][]{1997ApJS..112..221J}:

\begin{equation} \label{eq:basic}
\Delta \lambda \propto n_e^{2/3} \{ \frac{n^2}{n^2-4} \}^2 [n(n-1)+2]
\end{equation}

 It is interesting to understand the difference between density
 estimates using TB09$+$HM88 and S78 for the $t=2.2$~s of the F13 $\delta=3$
 simulations in Figure \ref{fig:f13}.
 Using the S78 theory gives a
 0.1 H$\gamma$ width of 16 \AA, which is a factor of nearly five lower than with
 the TB09$+$HM88 profiles.  Using the relation from the TB09$+$HM88
 broadening models in Figure \ref{fig:nel},
 a broadening of 16 \AA\ gives an electron density that is
 a factor of five to ten lower than the range in the RHD (RADYN)
 calculation.  The results of \citet{1997ApJS..112..221J} demonstrated
 a similar discrepancy between the S78
 treatment and the modified impact theory of \citet{Kepple1968} and \citet{Bengtson1970}.

According to Equation \ref{eq:basic}, the widths of higher order lines
should increase significantly (e.g., H$\alpha$ compared to H$\gamma$),
in the regime where thermal broadening is comparatively small.
However, the interline optical depth variation results in much less extreme
width differences \citep{Svestka1962, Svestka1963}.
In our models (and generally in the observations too) the broadening
is not strongly dependent on the Balmer transition because the much larger 
optical depth for H$\alpha$ near line center leads to a much smaller ($\sim100$x)
physical depth range over which the line is produced compared to H$\gamma$, thus giving a
larger value of the 0.1 width for H$\alpha$ than indicated by Equation \ref{eq:basic}.  We also
note that directly comparing to the width of the
S($\alpha$) profiles results in largely erroneous values of the
electron density;  due to the optical depth over the line, the values of
S($\alpha$) in the far
wings at S($\alpha) \sim10^{-4}$ are the important determinant in the
0.1 width of the line when the optical depth near line center and the
emissivity in the line wings are both
large, as in the Balmer line profiles of Vega.

In Figure \ref{fig:nel} the width of H$\gamma$ is shown from a 
new electron beam heating simulation with a flux of $5\times10^{12}$
erg cm$^{-2}$ s$^{-1}$ (5F12) that is calculated with the RADYN flare code,
using a high, low-energy cutoff of $E_c=150$ keV so that the heating occurs
deep in the atmosphere (see Appendix A).   The emergent continuum
spectrum has a small
Balmer jump ratio and the low order Balmer lines are broad and in emission with
a prominent
central reversal.   Like for Vega the H$\gamma$ line is formed over
a large range of electron
densities, given by the error bar range in Figure \ref{fig:nel}, which demonstrates that a line
width value does not always correspond to a unique electron density value;
this is also true for the F13 model atmospheres but for a narrower
range of electron density.  Interestingly, 
the broadening (75 \AA) in the H$\gamma$ emission line in the F13 model is
comparable to the broadening (nearly 70 \AA) in the H$\gamma$
absorption line in Vega, 
while there is an
order of magnitude difference in the maximum charge density over the
formation heights
of the 
line profile.  The H$\gamma$ line in the 5F12 $E_c=150$ keV
model forms over an electron density that is comparable to and larger
than the electron density that produces the line in Vega, but
the broadening is much larger in the Vega spectrum.

The model predictions of the highest order Balmer line in
absorption (at least H16) in the Vega spectrum (Figure \ref{fig:vega}
top) differ significantly from the
highest order Balmer line that is in 
emission (H9) in the F13 spectrum.
Therefore, the amount by which the higher 
order Balmer lines fade into the dissolved level continuum flux at wavelengths longer than the Balmer edge
can provide an additional constraint to break such a degeneracy and ambiguity in the
charge density-line width relation in flare
atmospheres.  In the 5F12 model, the highest order Balmer line that is
not completely dissolved is $\approx$H13, and the far wing of H$\gamma$
($\lambda_{\rm{rest}}+20$ \AA) forms over an electron density that is three times as
great as that in Vega (at the same wavelength) but a factor
of three lower than the F13 model (at the same wavelength).
This ordering by electron density is consistent with the highest order Balmer line 
in each model spectrum.
 \citet{Kowalski2015} discussed a
dissolved line decrement\footnote{Referred to as a ``L-Z decrement''
  in \citet{Kowalski2015}.}
using H11/H$\gamma$ that
can be used to quantify the degree by which the higher order lines
fade into continuum flux;  we intend to pursue this diagnostic in
a future work with NLTE flare models and the TB09$+$HM88 broadening.


\subsection{The Balmer Line Broadening from Many Flare ``Threads''} \label{sec:multithread}
F13 beam flux model atmospheres produce dense, downward
directed, heated
compressions (``chromospheric condensations'') and have been found to adequately explain the hot
($10^4$) blackbody-like, optical and NUV continuum
flux distribution of flares on active M dwarf stars
\citep{Kowalski2015, Kowalski2016}.  However, the new
broadening predictions show that these white-light emitting
compressions result in a charge
density ($\sim5\times10^{15}$ cm$^{-3}$) that
over-broadens the Balmer lines as in Figure \ref{fig:f13}
compared to observations in the literature.  If such dense
compressions are formed in dMe flares, then they would have to account for a
large fraction of the continuum flux and a small fraction
of the line flux, such that the spatially integrated flare
spectrum exhibits narrower lines than the TB09$+$HM88, F13 $t=2.2$~s prediction.

\citet{Kowalski2015} used a time-average of the F13 model atmosphere over its
evolution to more realistically simulate the prediction over an
exposure time.  Averaging over the F13 simulation is equivalent to an 
\emph{instantaneous} measurement of a spatially unresolved observation
of many F13 flare loops 
sequentially being initiated at regular, short time intervals
corresponding to different locations on the star.
 This prescription for representing the flare flux from many flare
 ``threads'', or kernels, is similar to  the
``multithread'' modeling that has been successful in reproducing
spatially unresolved 
solar flare light curves in GOES soft X-rays \citet{Warren2006}.  A method for multithread modeling
with RADYN heating models lasting $\sim15-20$~s was further developed
in \citet{Fatima2016} and has also been employed in \cite{Reep2017}.   In multithread modeling of solar flares, the
flux of each thread changes over the evolution of the flare
\citep{Warren2006} such that impulsive phase threads have a higher
flux than gradual phase threads.  For
simplicity, we use the F13 model to represent all heated threads (kernels).

We simulate a multithread prediction for the line broadening from the
F13 simulations in \citet{Kowalski2015} and \citet{Kowalski2016}.
We use the snapshots from Figure \ref{fig:nel} at $t=0.2$, $1.2$, 2.2,
and 4.0~s (the electron beam heating is turned off at
2.3~s and allowed to relax until $t=5$~s) to sample the 
range of atmospheric conditions and brightness values achieved in the
models.  The assumption of statistical equilibrium (without the
  occupational probability formalism of \citet{HHL}) was justified in
Sections \ref{sec:vega} and \ref{sec:stat} but may not be appropriate within the earliest
tenths of seconds after
the beam heating starts.  The Balmer lines are broadest at $t=2.2$~s
when the chromospheric condensation has achieved the maximum electron
density at $T\sim12,000-13,000$ K, thus averaging over earlier and
later times decreases the line broadening in the multithread models.  For
the F13 dpl multithread (0-5~s average) model, the 0.1 width of
H$\gamma$ is $\sim50$ \AA, a factor of 2 narrower than at $t=2.2$~s;
for the F13 $\delta=3$ multithread (0-5~s average) model, the 0.1 width is 40 \AA\ (compared to 75 \AA\ at
2.2~s; Figure \ref{fig:f13}).  The 0.1 width of the F13 $\delta=3$ is near the upper range of 
observed values in the early impulsive phase but are still quite
high.  Whereas the S78
predictions from the F13 model at 2.2~s for the continuum flux
ratios\footnote{For example, C3615/C4170 is the ratio of the continuum
  flux averaged in the 30\AA\ window around $\lambda=3615$ \AA\ to the continuum flux
  averaged in a 30\AA\ window around $\lambda=4170$ \AA.}
(C3615/C4170, C4170/C6010), the line-to-continuum ratio values of H$\gamma$/C4170 ($=10-20$), and
broadening are generally consistent with peak phase spectra of dMe
flares \citep{Kowalski2015, Kowalski2016}, the time-averaged values with TB09$+$HM88 broadening are more
consistent with the early impulsive rise phase of some dMe flares in
the literature, as concluded in \citet{Kowalski2015} for the
time-average continuum properties of the F13 dpl model.  For example, the time-resolved impulsive phase data
of the large IF3 flare event from \citet{Kowalski2013}
exhibit values of (C3615/C4170, C4170/C6010, H$\gamma$/C4170, 0.1
width H$\gamma$)$=$(2.2, 1.8, 40, 48 \AA) in the early-to-mid rise phase
(S\#27, cf Figure 30 of \citet{Kowalski2013}) that are very similar to these
quantities (2.1, 1.8, 45, 45 \AA, respectively) from the F13
$\delta=3$ multithread model (Table \ref{table:decrement}).

\subsection{Balmer Decrements\label{sec:decrement}}
Balmer decrements are defined as the ratio of the excess (i.e.,
background subtracted) flux in Balmer lines to that of the H$\gamma$ line
and are also used to constrain the charge density variations in flares
\citep{HP91, Jev98, Garcia2002}.
The observed decrements in dMe flares are typically 
H$\beta$/H$\gamma \sim 1.0-1.2$ and for H$\delta$/H$\gamma \sim
0.75-1.05$ (Table 4.20 of \citet{Kowalski2012}, Table 1 of \citet{Allred2006}).  
Decrements are similarly used in solar flare studies as well, but the
lack of broad wavelength coverage spectra make the measurements rare
in the modern era
\citep{1997ApJS..112..221J, KCF15}.  Since Balmer decrements have been widely used, it is informative to
study how they may vary in response to electric pressure
broadening with the new TB09$+$HM88 profiles.

In Table \ref{table:decrement} we show the decrements
for the two F13
models at $t=2.2$~s and the 0-5~s average (multithread) model (Section
\ref{sec:multithread}) calculated with the TB09$+$HM88
broadening.  Although the line flux increases by $\sim2.5$x with TB09$+$HM88 compared to
the line flux calculated with the S78 broadening, 
the line flux ratios do not change appreciably between
the two methods.  
The H$\beta$/H$\gamma$ and H$\alpha$/H$\gamma$ line flux ratios are much smaller in the models than typical 
values from observations, but the
H$\delta$/H$\gamma$ ratios are in general agreement.  Such small values of
H$\alpha$ and H$\beta$ to H$\gamma$ indicate a ``reverse decrement''
and are only observed to be as low as $\sim0.8$  (for H$\alpha$;
Figure 4.22 of \citet{Kowalski2012}).  
The H$\alpha$/H$\gamma$ line flux ratios are very low in the F13 models because the
optical depth necessary to produce the hot blackbody-like continuum
shape results in a very large H$\alpha$ optical depth and radiation
thus escapes from a much smaller physical depth range of the atmosphere across this line compared to H$\gamma$.
We've also computed the line flux ratio at $t=0$~s in our M dwarf
model (row 1), which is in good agreement with observations
\citep[][see also Table 2.8 of \cite{Kowalski2013}]{Bochanski2007} but with a
significantly larger
line flux ratio for H$\beta$ and slightly larger ratio for H$\alpha$ than in the observations. The discrepancy could result from the difference 
between spatially resolved and averaged observations, and 3D effects
\citep{Walkowicz2008, Uitenbroek2011, Leenaarts2012, Wedemeyer2015}.  As in the flare simulations, the
H$\delta$/H$\gamma$ line flux ratio is consistent with the
observations.  

It is interesting to compare the values in Table \ref{table:decrement} to the method of \citet{Drake1980}, which
presented the Balmer line decrements for a large parameter space of
electron density, temperature, and optical depth for a homogeneous slab of
hydrogen.  In addition, approximations were made using the escape
probability from the line wing and the optical depth variation from
line to line.  In their Appendix B, they present the decrements for
$n_e=10^{15}$ cm$^{-3}$, which is the highest density that they
considered:  the H$\alpha$/H$\gamma$, H$\beta$/H$\gamma$, and H$\delta$/H$\gamma$ decrements are 1.8,
1.2, and 0.85, respectively.  The values for their limiting optically
thick case, on the other hand, show reverse decrements for H$\alpha$ and
H$\beta$ but are not as extreme as for the F13 model at $t=2.2$~s, likely
due to the factor of 5 larger electron density in the F13 model.  The results from
\cite{Drake1980} are qualitatively similar to ours for flare
atmospheres that are so dense that they are approximately homogeneous
in electron density (e.g., at $t=2.2$~s of the F13 dpl model, the
contribution function-weighted electron
density varies from only $4-6\times10^{15}$ cm$^{-3}$ over the H$\gamma$ line), but would not be
applicable to the flare atmospheres that exhibit a large range of
electron density over the line formation, such as the high, low-energy
cutoff (5F12, $E_c=150$ keV) flare model in Figure \ref{fig:nel} with
a range of nearly hundred in electron density.

\floattable
\begin{deluxetable}{lcccccccc}
\tabletypesize{\scriptsize}
\tablewidth{0pt}
\tablecaption{Balmer Decrements and H$\gamma$ Broadening with TB09$+$HM88}
\tablehead{
\colhead{Model} &
\colhead{H$\alpha$} &
\colhead{H$\beta$} &
\colhead{H$\gamma$} & 
\colhead{H$\delta$ }&
\colhead{0.1 width H$\gamma$ [\AA]}&
\colhead{H$\gamma$/C4170} &
\colhead{C3615/C4170} &
\colhead{C4170/C6010} }
\startdata
pre-flare* & 5.9 & 2.7 & 1.0 & 0.58 & \nodata & \nodata & \nodata &\nodata \\
F13 2.2s dpl & 0.3 & 0.7 & 1.0 & 0.8 & 100 (100) & 45 & 1.8 & 2.1 \\
F13 dpl multithread (0-5~s ave) & 0.4 & 0.8 & 1.0 & 0.9 & 52 (60) & 74 & 2.6 & 1.6\\
F13 2.2s $\delta=3$ & 0.3 & 0.8 & 1.0 & 0.9 & 75 (78) & 31 & 1.7 & 2.3  \\
F13 $\delta=3$ multithread (0-5~s ave) & 0.4 & 0.9 & 1.0 & 1.0 & 40 (45) &
45 &
2.1 & 1.8 \\
\enddata 
\tablecomments{*The pre-flare atmosphere is the most up-to-date
  version of the pre-flare state in our modeling, which is described in
  Appendix A.  The continuum flux ratios C3615/C4170 and C4170/C6010,
  and the line-to-continuum flux
  ratio (H$\gamma$/C4170) 
are also given. The values in parentheses for the H$\gamma$ widths are
calculated with an instrumental convolution of $R=450$.}
\label{table:decrement}

\end{deluxetable}



\section{The Line Broadening and Flux Decrements in the YZ CMi Megaflare}\label{sec:megaflare}

\begin{figure}[htbp]
\begin{center}
\includegraphics[scale=0.35]{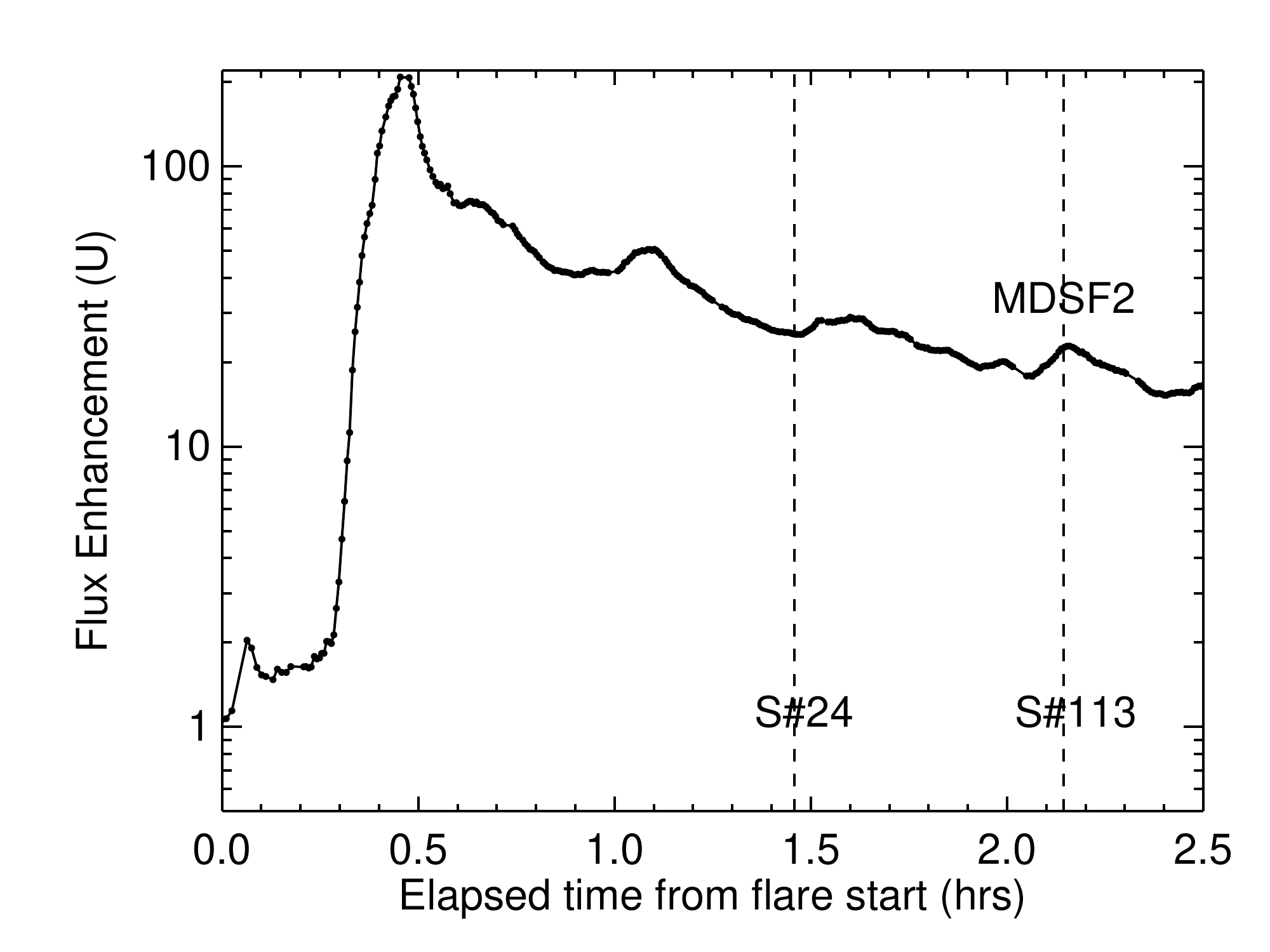}
\caption{U-band light curve of the YZ CMi Megaflare event.  The times
  of the gradual decay phase spectrum (S\#24) and the peak spectrum
  (S\#113) of the secondary flare MDSF2 are indicated.  Note that
  several secondary flares also occurred before S\#24.  }
\label{fig:megaflare_lc}
\end{center}
\end{figure}


In this Section, we model the Balmer decrements and broadening 
during the Megaflare on the dM4.5e star YZ CMi \citep{Kowalski2010}, which has
high-time resolution, flux-calibrated NUV and optical spectra.  The
$U$-band light curve from \cite{Kowalski2010} is shown in Figure
\ref{fig:megaflare_lc}.  The
Megaflare ($E_U>10^{34}$ erg) consisted of a
series of high-energy ($E_U\sim10^{32}$ erg) secondary flares after
the big first flare event at $t=0.45 $ hrs in Figure
\ref{fig:megaflare_lc}.  

\begin{figure}[htbp]
\begin{center}
\includegraphics[scale=0.65]{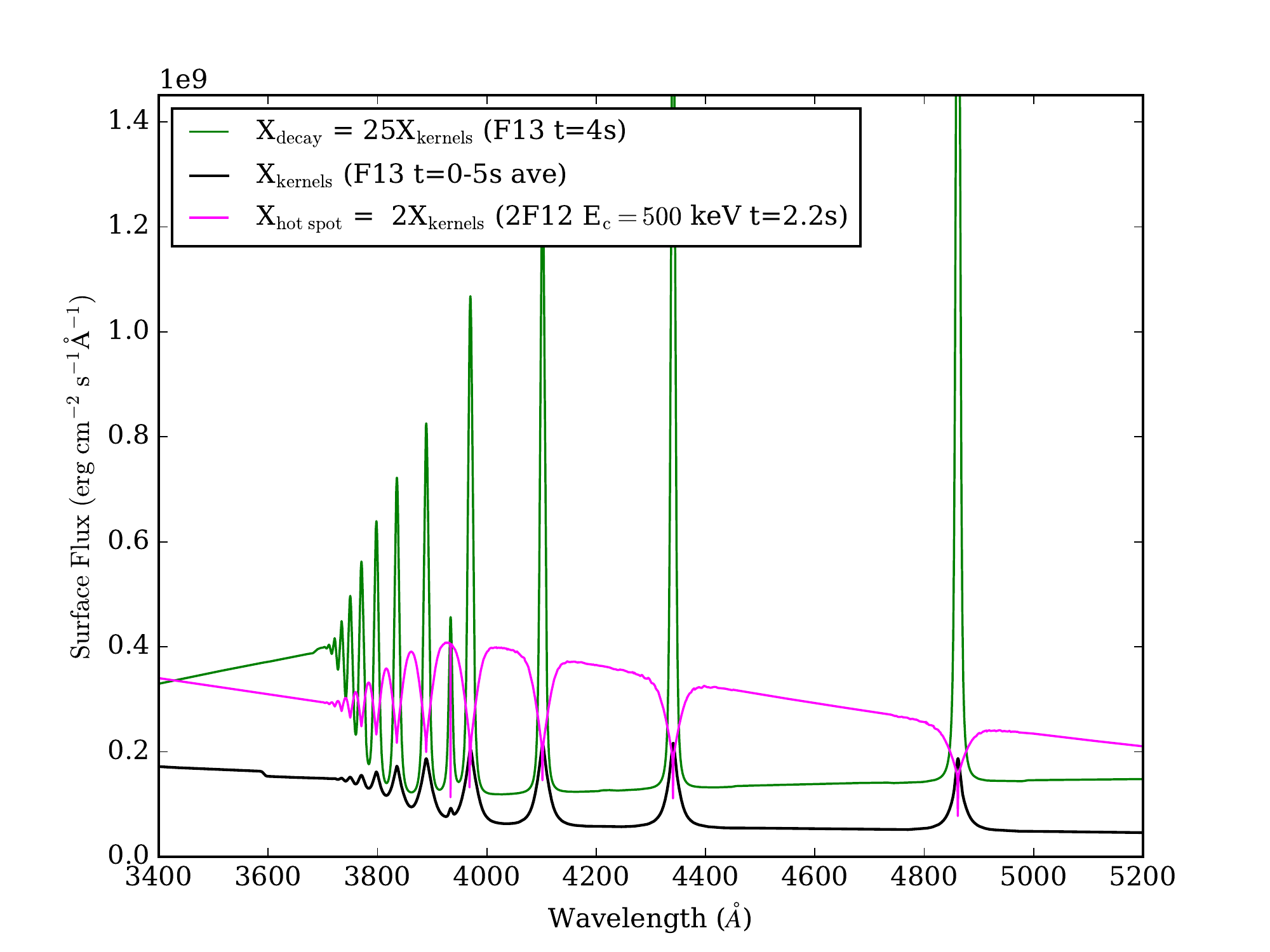}
\caption{The surface flux spectra for each of the components
  ($F_{decay}$, $F_{kernels}$, $F_{hotspot}$) multiplied by the respective
  filling factor, $X$.  The models are superposed with these filling
  factors to produce the model spectra in Figure
  \ref{fig:megaflare}.  The relative filling factors are indicated by
  the values of $X$ in the legend.  In Figure \ref{fig:megaflare} (top),
$X_{kernels}=0.0009$ and in Figure \ref{fig:megaflare} (bottom
panel), $X_{kernels}=0.00047$.  These spectra were calculated with RH
and the TB09$+$HM88 profiles.  }
\label{fig:megaflare_models}
\end{center}
\end{figure}

During the rise and peak of the secondary flare MDSF2 in the decay
phase of the Megaflare,
a blackbody fit to the blue continuum ($\lambda=4000-4800$ \AA) indicates an increasing color temperature while the 
line fluxes decrease for the (quiescent-subtracted) flare spectrum.  Subtracting the 
decay phase spectrum just before the start of MDSF2 reveals the spectrum similar to an A
star \citep[Section 6.3 and Figure 25 of][]{Kowalski2013};  the
veiling from this spectrum causes the Balmer flux decrement of the total flare
flux to tend towards an A0-star
decrement and the 0.1 widths of the emission lines in the total flare
spectrum to decrease below the resolution limit (13 \AA). This flare flux
component has been
modeled as a phenomenological hot spot with $T_{\rm{max}}=20,000$ K in
the photosphere \citep{Kowalski2011IAUS}.  We have successfully
reproduced a similar spectrum with the RADYN code using a very high,
low-energy cutoff of the electron
beam, $E_c=500$ keV, a moderately large flux of $2\times10^{12}$
erg cm$^{-2}$ s$^{-1}$, and a very soft spectrum, $\delta=7$.
Accurate treatment of the energy loss from this electron beam
requires the fully relativistic Fokker-Planck solution that is currently
employed in the RADYN flare code \citep{Allred2015}\footnote{Whereas the F13
models here were calculated with the non-relativistic prescription in
\citet{Emslie1978}.}. 
As in the F13 models, the 
electron beam energy flux at the top of the atmosphere is kept constant for 2.3~s and the atmosphere is allowed to
relax after 2.3~s.  Shocks do not develop in this model, and the atmosphere
can relax to much longer times than the F13, and the simulation takes
much less computation time.  We use the RH code with the
TB09$+$HM88 profiles for Balmer line broadening and level dissolution 
and calculate the emergent flux spectrum at $t=2.2$~s.  The
model spectrum is shown as the pink spectrum in Figure \ref{fig:megaflare_models}. 

The Balmer lines and Balmer continuum are in
absorption in the emergent flux spectrum, similar to the A0 star Vega; 
we thus use this 
beam-heated atmosphere to represent the newly-formed flux during 
the secondary flare MDSF2.   The properties of this model atmosphere
are discussed in Appendix A.  In summary, a temperature maximum peaking at nearly
$T\sim13,000$ K is formed in the very low chromosphere where the
temperature is 4000 K before the flare heating starts.  We 
interpret the deep heating in this model as a  ``hotspot''
as in the  static, phenomenological model of \citet{Kowalski2011IAUS}.
With the new RHD model, we have deduced that the hot spot can form much
higher (log $m/ \rm{g\ cm}^{-2} = -2.05$ to $-1.25$) than in the
phenomenological static model (log $m/ \rm{g\ cm}^{-2}=0.5$).

We compare the new RHD models with new (TB09$+$HM88) broadening to spectral observations of the flare-only (quiescent subtracted)
flux at two times in the YZ CMi Megaflare: spectrum S\#24 consisting mostly of decay phase flux at the tail end of
a secondary flare and spectrum S\#113 at
the peak of the secondary flare MDSF2.
The times of S\#24, S\#113, and MDSF2 are
indicated in Figure \ref{fig:megaflare_lc}.
The observed properties of the YZ CMi Megaflare are presented in \citet{Kowalski2013} and are summarized
in Table \ref{table:megaflare} (rows 1-2) for S\#24 and S\#113.  We also include continuum flux ratios
(C3615/C4170 and C4170/C6010)
here to compare to the model predictions of the continuum shape.  
Following \citet{Hawley2003} and \citet{Osten2016}, we model the
flare-only
flux 
observed at Earth as:

\begin{equation} \label{eq:add}
F_{flare-only, Earth}=[F_{kernels}X_{kernels} + F_{decay}X_{decay} +
F_{hotspot}X_{hotspot}] \frac{R_{\rm{star}}^2}{d^2}
\end{equation}

\noindent where $d$ is the distance to YZ CMi and
$R_{\rm{star}}=0.3R_{\rm{Sun}}$, $F$ is a surface flux spectrum of a
model component and $X$ is the filling factor, which is the fractional area of the
visible stellar hemisphere that is emitting at each value of $F$
\citep{Hawley2003}.   The three model
component spectra ($F_{kernels}, F_{decay}, F_{hotspot}$) superposed to
produce the model flare-only\footnote{The precise comparison of the model predictions to
  the observed flare-only flux requires subtracting the pre-flare
  surface flux spectrum scaled by the filling factor of each flare
  surface flux component \citep[Equation 3 of][]{Kowalski2016}; here, the flare surface
  flux spectra are much greater than the pre-flare surface flux
  spectrum and thus Equation \ref{eq:add} is a sufficient comparison to
  the observed flare-only flux.} flux at Earth are shown in
Figure \ref{fig:megaflare_models}.  
 $F_{kernels}$ is the F13 dpl
multithread model (0-5~s average; Table \ref{table:decrement}),
$F_{decay}$ is the F13 dpl model at $t=4$~s, and $F_{hotspot}$ is the 2F12
$E_c=500$ keV model at $t=2.2$~s.   Each flux component has been calculated
using RH with the TB09$+$HM88 profiles.   The phenomenology of ``kernels'',
``decay'', and ``hotspot'' is discussed in Section \ref{sec:solarstellar}.   

Following the analysis of the DG CVn superflare presented in \citet{Osten2016}, we assume $X_{decay}=25X_{kernels}$
and $X_{hotspot} =0$ for S\#24 and solve for $X_{kernels}$ using the
flare-only, specific
continuum flux centered at $\lambda=4170$ \AA\ and $\lambda=4785$ \AA\ observed
at Earth.  For S\#113, we assume
$X_{decay}=25X_{kernels}$ and find that $X_{hotspot}=2X_{kernels}$ by 
fitting to the Balmer jump ratio (C3615/C4170 $=1.5$) in the observation;  then we
solve for $X_{kernels}$ using the flare-only, specific continuum flux centered at $\lambda = 4170$
\AA\ and $\lambda=4785$ \AA\ observed at Earth.  
The
value of $X_{kernels}$ is 0.0009 for S\#24 and $0.00047$
for S\#113.  We refer to the model for S\#24 as 
``the DG CVn Superflare F13 multithread model'' and the model for
S\#113 as ``the DG CVn Superflare F13 multithread model $+$ Hot
Spot''.   

\begin{figure}[htbp]
\begin{center}
\includegraphics[scale=0.65]{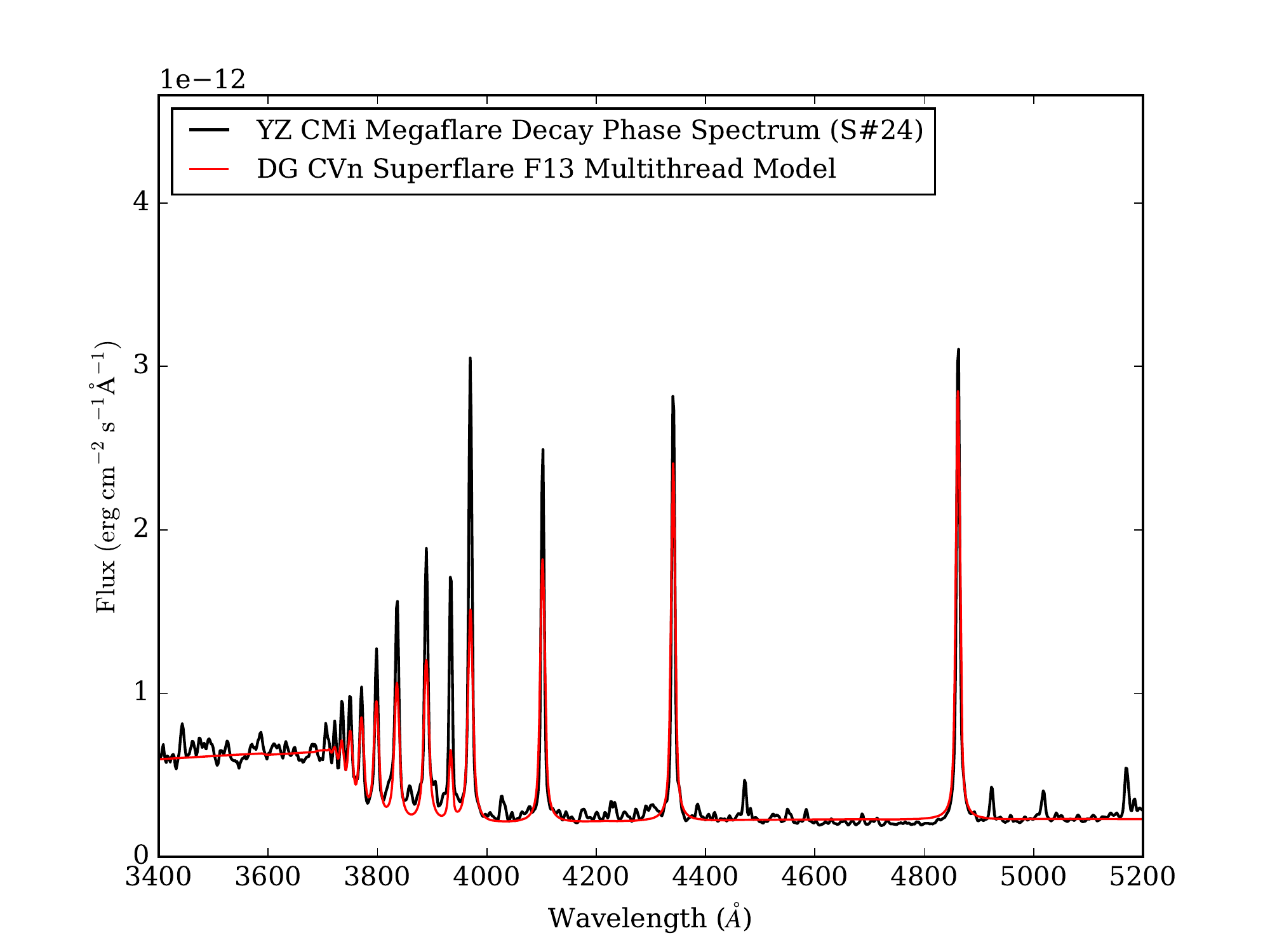}
\includegraphics[scale=0.65]{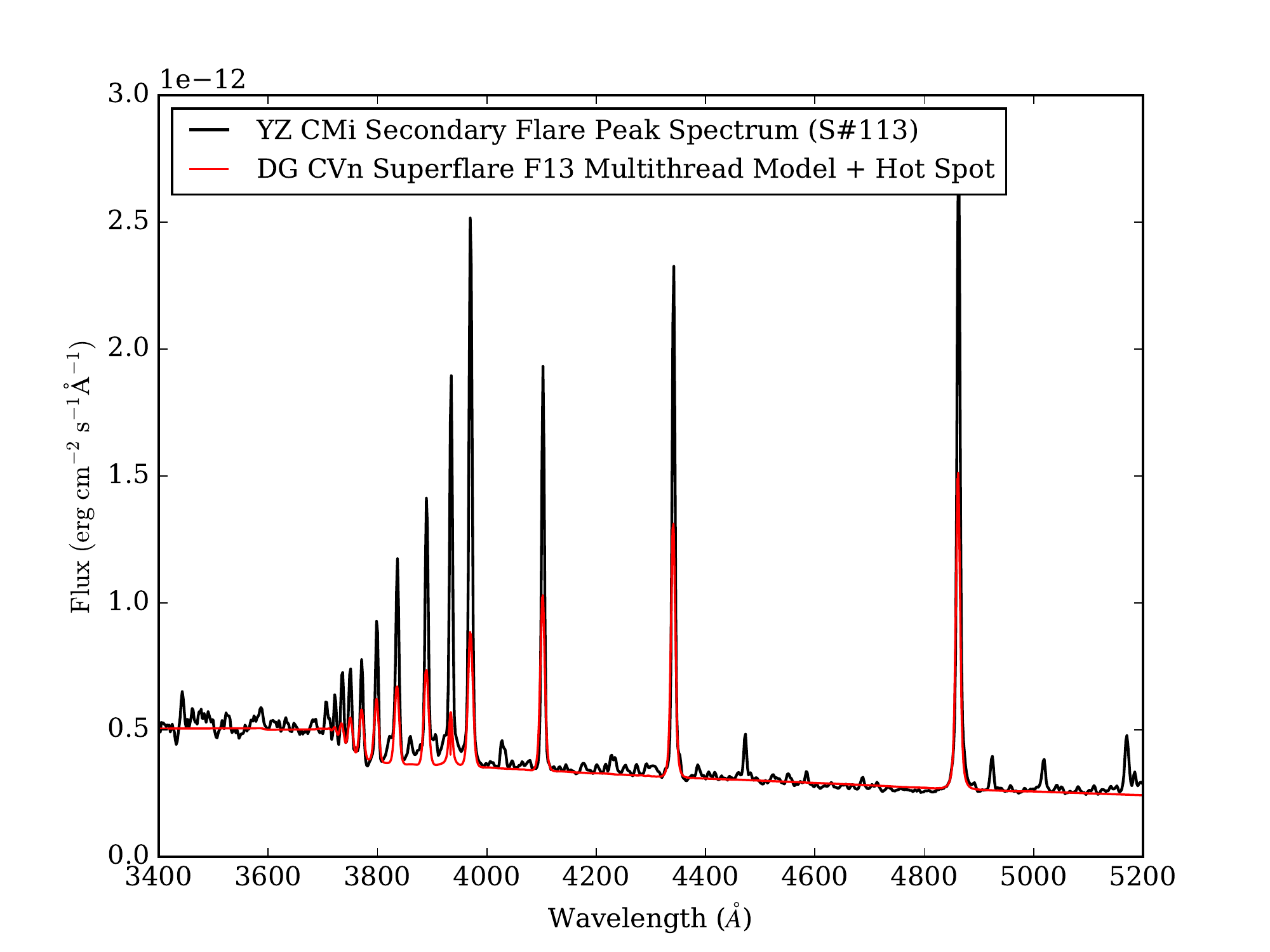}
\caption{(top) Model (red) of the flare-only flux for the spectral observation S\#24 (black) in the decay phase of the YZ CMi
  Megaflare. The observed and model quantities are summarized in the
  1st and 3rd rows of
  Table \ref{table:megaflare}, respectively.  (bottom) Model (red) of
  the flare-only flux for the spectral observation S\#113 (black) of 
  the spectrum at the peak of the secondary flare MDSF2 in the YZ CMi Megaflare.
  The observed and model quantities are summarized in rows 2 and 4 of Table
  \ref{table:megaflare}, respectively.  The model spectra have been convolved with a
  Gaussian with FWHM $=$ 6 \AA\ for direct comparison to the
  observations.  The changes in the observed continuum shape and 
  line broadening and fluxes are qualitatively
  reproduced by the models and are in general quantitative agreement,
  however more model flux is needed in the higher order Balmer lines and Ca II while the
  Balmer lines are slightly over-broadened.  The
  component model surface flux spectra are shown in Figure \ref{fig:megaflare_models}. }
\label{fig:megaflare}
\end{center}

\end{figure}

The results of the model comparison to the observation of S\#24 are
shown in the top panel of Figure \ref{fig:megaflare}; the Balmer line
flux decrements and broadening, and continuum flux ratios of the model
are shown in
the third row of Table \ref{table:megaflare}.   
 We find that the DG CVn Superflare F13  multithread model improves the
 model broadening and
decrements of the Balmer lines that were found in Section
\ref{sec:multithread}.   The widths and flux decrements
are in striking agreement with the decay phase spectrum (S\#24) of the Megaflare, as are the
continuum flux ratios.  From Table
\ref{table:decrement} it is clear that neither the F13 multithread model (0-5~s average) nor the instantaneous F13 model at $t=2.2$~s
alone account for the continuum and line properties in the YZ CMi
Megaflare decay phase (S\#24); an
additional filling factor of the instantaneous F13 model prediction at
$t=4$~s is necessary to decrease the broadening and increase the
H$\beta$/H$\gamma$ decrement in the flare-only model spectrum.  The
continuum flux ratio C4170/C6010 also becomes lower, in agreement
with the observations.  The H$\gamma$/C4170 value is higher 
than in the observations but still in general agreement.  Apparently,
the same two-component multithread model in the decay phase of an
$E\sim10^{36}$ erg secondary flare event \citep[F2;][]{Osten2016} during the DG CVn Superflare can also
adequately explain the Balmer line \emph{and} continuum flux properties in
the decay phase (S\#24) of the secondary flare events (e.g., at $t=1.1$ hrs in
Figure \ref{fig:megaflare_lc}) in the YZ CMi Megaflare.  For a comparison 
of the continuum flux properties in the DG CVn Superflare and in the YZ CMi
Megaflare,
we refer the reader to \citet{Osten2016}.


For S\#113, we use the 2F12 $E_c=500$ keV (hotspot) model to represent
the newly formed flare emission at the peak of the secondary flare
MDSF2 and add this model component to the two-component DG CVn
Superflare multithread model (Equation \ref{eq:add}).
The resulting line and continuum properties for the superposed flare
flux model are shown in the last row of Table \ref{table:megaflare}, and 
the model spectrum and the observation of S\#113 are shown
in the bottom panel of Figure \ref{fig:megaflare}.  The superposition
of the 2F12 $E_c=500$ keV (hotspot) model with the DG CVn
Superflare multithread model explains several observed
changes in the flare-only spectra from S\#24 (top panel of Figure
\ref{fig:megaflare}) to S\#113 (bottom panel of Figure \ref{fig:megaflare}).  
The model H$\beta$/H$\gamma$ ratio increases and
the H$\delta$/H$\gamma$ ratio decreases, as observed.  
The line width of H$\gamma$ decreases due to the superposition of the broad absorption
wings from the high, low-energy cutoff model ($E_c=500$ keV) and the emission lines
from the DG CVn Superflare F13 multithread model;  the H$\gamma$/C4170 ratio
decreases from 120 to 30, qualitatively similar to how the observations decrease from 90 to 40, and the red continuum (C4170/C6010) 
becomes bluer (hotter) as in the observations.  The H$\alpha$
line was saturated for these spectra, but we give the model values
to show that the values of 0.8-1.0 are consistent with observed values of
the reverse decrement in other flares.  

In the models in Figure \ref{fig:megaflare}, the opacity effects from level dissolution are also
included, showing that the spectral properties at $\lambda=3646-3800$ \AA\ are adequately
reproduced by the merging of the Balmer line wings and
the continuum flux longward of the Balmer edge\footnote{Note the
  feature near 3704 \AA\ in the observation is a He I line which is
  blended with the H15 and H16 Balmer lines.}.  To match the models
precisely to the observations, more flux and less broadening in the Balmer lines
 and more flux in the Ca
II K line are needed in the models without largely affecting the
continuum flux
distribution (the H$\gamma$/C4170 in the model
for S\#24 is higher than the observations, but appears lower in flux
because of the slightly larger broadening).  Additionally, a smaller amount of level dissolution for
the higher order Balmer lines is required in the total model spectra
of the flare-only flux.  In Section \ref{sec:gradual}, we speculate on future
modeling improvements to explain these discrepancies.

\floattable
\begin{deluxetable}{cccccccc}
\tabletypesize{\scriptsize}
\tablewidth{8in}
\tablecaption{Comparison to the YZ CMi Megaflare}
\tablehead{
\colhead{Observation (S\#)} &
\colhead{H$\alpha$/H$\gamma$} &
\colhead{H$\beta$/H$\gamma$} &
\colhead{H$\delta$/H$\gamma$} &
\colhead{H$\gamma$ 0.1 width [\AA]} &
\colhead{H$\gamma$/C4170} &
\colhead{C3615/C4170} &
\colhead{C4170/C6010}  }
\startdata
decay phase observation* (S\#24) & -- & 1.26 & 0.86 & 14.6 & 87$\pm11$ & 2.7 & 1.1 \\
secondary flare observation* (S\#113; MDSF2) & -- &  1.50 & 0.75 &12.6 & 42$\pm2$ & 1.5 & 1.6 \\
\hline
Model & & & & & & & \\
\hline
DG CVn Superflare F13 multithread model** & 0.83 & 1.18 & 0.8 & 14.4 (21) & 120 & 2.9 & 1.0 \\
DG CVn Superflare F13 multithread model $+$ hot spot** &  1.0 & 1.3 & 0.7 & 12.6 (18) & 30 & 1.5 & 1.7 \\
\enddata 
\tablecomments{*These spectra include the flare-only flux with the
  quiescent spectrum subtracted (see \cite{Kowalski2013}). S\#24 is
  shown in Figure \ref{fig:megaflare} (top) and S\#113 is shown in
  Figure \ref{fig:megaflare} (bottom).  **``The DG CVn Superflare F13
  multithread model'' is the F13 dpl model averaged over 0-5~s with a filling factor of
$X_{kernels} = 0.0009$ added to the F13 dpl model at $t=4$~s with a filling
factor of $25X_{kernels}$;  this model is shown in the top panel of
Figure \ref{fig:megaflare}.  The ``DG CVn Superflare F13 multithread
model $+$ hot spot'' is the F13 dpl model averaged over 0-5~s with a filling factor of
$X_{kernels} = 0.00047$ added to the F13 dpl model at $t=4$~s with a filling
factor of $25X_{kernels}$ plus the spectrum at $t=2.2$~s from the 2F12
$E_c=500$ keV simulation with a filling factor of $2X_{kernels}$; this
total model spectrum is shown in the bottom panel of Figure \ref{fig:megaflare}.
The 0.1 width values in parentheses were calculated with a Gaussian instrumental convolution with R $=$ 670
  as in the observations in Figure \ref{fig:megaflare}. }
\label{table:megaflare}
\end{deluxetable}

\subsection{Interpretation:  Extending the Solar Flare Analogy to
  Megaflares}  \label{sec:solarstellar}

 Following \citet{Kowalski2012SoPh}, we interpret the surface flux
 spectra components (Figure \ref{fig:megaflare_models}) in the S\#24
 and S\#113 spectral observations of the YZ CMi
Megaflare using analogous phenomenology from high spatial resolution 
data of two-ribbon solar flares.

 \begin{itemize}

\item 
$F_{kernels}$:  The F13 multithread model spectrum (0-5~s average of
the F13 dpl model) represents 
the spreading ribbons of the Megaflare.  As the ribbons spread,
new flare loops form, and their initiation is staggered in
time so that the spatially integrated flux from this area is
equivalent to the $0-5$~s time-average of the F13.  The area decreases
over time ($X_{kernels}$ decrease from 0.0009 at the time of S\#24 to 0.0005 at
the time of S\#113); the beam
flux probably also decreases over the gradual phase \citep{Warren2006} as lower
field strengths reconnect higher in the corona.  We note that newly heated flare loops are necessary to explain
the temporally extended gradual phase of the soft X-ray flux  in spatially
unresolved GOES light curves 
of solar flares \citep{Warren2006}.

The spreading ribbons may correspond to the regions that produce the
end of the large secondary flare (e.g., at $t=0.6$ hrs or $t=1.1$ hrs in Figure \ref{fig:megaflare_lc}, before S\#24) that peaked prior to the
spectral observations, and/or they may be the
continuation of the spreading ribbons from the big first flare event
with $\Delta U=-5.8$ mag at $t=0.45$ hrs in Figure \ref{fig:megaflare_lc}.
 Recent high spatial resolution images in the H$\alpha$
red wing ($+0.8$\AA) with the New Solar Telescope (NST)
have shown that spreading H$\alpha$ ribbons are composed of many fine-scale
kernels with a size of $\sim100$ km \citep{Sharykin2014, Jing2016};  these kernels are
located at the leading edge \citep{Isobe2007} of
the ribbons and are often associated with red-wing asymmetries in
H$\alpha$.  Other
chromospheric flare lines, such as Mg II and Fe II, also exhibit red-wing
asymmetries.  \citet{Kowalski2016IRIS}
reproduces the general properties of the red-wing
asymmetries in Fe II lines in the brightest flare footpoints
constituting the two flare ribbons that 
spread apart in the impulsive phase of the 2014 Mar 29 X1 solar
flare.  The RHD model that produces dense, heated chromospheric
compressions that are consistent with the spectral observations employed a relatively high
electron beam flux (5F11).  
The F13 model evolution consists of 
a dense, heated chromospheric condensation but with higher 
temperature, density, and continuum optical depth compared to the
chromospheric condensation properties in
the 5F11 model for solar flares \citep[see Appendix C of][]{Kowalski2016IRIS}.  Thus, the kernels of the spreading
ribbons may consist of many $\approx$F13 beams in dMe flares
and lower fluxes, $\approx$5F11
beams, in solar flares.

\item $F_{decay}$:  We use 
the F13 $t=4$~s flux to represent the wake of the spreading flare
ribbons $F_{kernels}$, resulting in long-lasting (temporally extended)
decay phase flux in the Megaflare that persists for hours and is
present through the spectral observations S\#24 and S\#113.  Note,
that the F13 multithread model (0-5~s average) also consists of this
decay phase
snapshot ($t=4$~s), but its contribution in the 0-5~s average represents the
rapid decay phase of a kernel.  
The gradually decaying flux may be analogous to temporally extended decay flux in the
``wake'' of separating ribbons in solar flares, such as observed after
the peak of the
NUV ($\lambda2826$) continuum-emitting kernels in the
2014 March 29 X1 solar flare \citep{Heinzel2014, Kowalski2016IRIS} and
the temporally extended gradual phase component of optical flare
kernels on the Sun
\citep{Kawate2016}.  
We assume that the first large event in the Megaflare
(or any of the secondary large events peaking before S\#24)
produces a wake of persistent, decaying flux behind the spreading ribbons. 
In our modeling, this flux component is not self consistent in the timing 
relative to one of these peaks in the Megaflare;  the NUV continuum exponential
decay constant after $t=2.3$~s in the F13 model is $\tau \sim 0.6$~s, meaning that the
pre-flare continuum level would be reached several seconds after the beam
heating ends (the F13 models are very computationally demanding,
preventing us from currently following the evolution for hours after the
beam heating).
In Section \ref{sec:megaflare}, we found that this  flux component is necessary in order to
 lower the Balmer line decrements and broadening in the
model to the observed range of values in S\#24.  In Section \ref{sec:conc} we speculate on 
directions of future work to better model and constrain the origin of
the temporally extended decaying flux component.  

We use the high spatial resolution data in the H$\alpha$ red wing from the
DST/IBIS presented in \citet{KCF15} to estimate the ratio of newly
formed kernel area in the spreading flare ribbons to the the area of the intensity in the wake of the
ribbons.  The IBIS data have a cadence of 20~s and
covered several peaks in the X-ray impulsive phase of a two-ribbon,
long-duration C1 solar flare, SOL2011-08-18T15:15.   For each frame, we
measure the area of bright flare intensity that is not bright in the previous
frame.  We only consider the umbral ribbon, which spreads very
slowly across the umbra;  most of the apparent kernel motion is parallel along the
ribbon.  The area of the newly
formed H$\alpha$ kernels is only a small fraction (0.1-0.2) of the total
H$\alpha$ kernel area emitting above the threshold count rate.
Thus, it is reasonable
to assume that the decay flux (``ribbon wake''; F13 $t=4$~s)
and the newly heated kernels (F13 $t=0-5$~s average) could reasonably exist
in the areal ratio that we employ (1/25) for the DG CVn Superflare and the YZ
CMi Megaflare events.

\item $F_{hotspots}=F_{new\ ribbons}$:
During the rise and peak of the secondary flare MDSF2, new flare
ribbons consisting of new kernels develop.  From our RHD modeling of
the new flare flux in the secondary flare
with the $2\times10^{12}$ erg cm$^{-2}$ s$^{-1}$ $E_c=500$ keV beam
(Appendix A),
we infer that the heating scenario in the secondary flare
is strikingly different than the heating
in the F13 models with $E_c=37$ keV, where the electron beams lose their energy in the
upper chromosphere.  In our terminology, \emph{hotspots} form 
 in the lower chromosphere, which is necessary to 
produce the Vega-like flare spectrum, and \emph{kernels}
produce chromospheric condensations in the upper chromosphere.  $ F_{new\ ribbons}$
may produce F13-type heating in addition to hotspot heating.  In
MDSF2, we infer that the dominant heating goes into forming the hotspot.  The filling factor of the
hotspot, $X_{hotspot}$, is 0.001, which is remarkably similar to
the filling factor inferred from a blackbody fit to the
blue continuum with
$T=10,000$ K \citep{Kowalski2010}.  The filling factor of $F_{decay}$
is 12.5x greater than this, which was also inferred in \citet{Kowalski2010} by fitting an optically
thin Balmer continuum \citep{Allred2006} to the flux at $\lambda < 3646$ \AA\ to
represent the decaying H$\alpha$ ribbons from the main peak.  
In the three-component phenomenological model of
\citet{Kowalski2012SoPh}, the $F_{kernels}$ component was modeled
as a second (cooler)
hotspot in the photosphere.


\end{itemize}

 In the superflare from DG CVn \citep{Osten2016}, $F_{kernels}$ was used
to represent the newly heated loops after the peak of the high energy
secondary flare 
event (referred to as F2).  
The energetic secondary
flare F2 had an impulsive phase timescale that was much longer
than the ``big first flare'' (BFF) event of comparable energy an hour earlier.  Thus,
different heating scenarios may dominate through the BFF
and F2 events in the DG CVn Superflare.  A significant detection of
continuum flux with a cool color temperature was obtained from the UVW2/V-band and V-band/R-band flux ratios,
but the only simultaneous observations to constrain this component
occurred after the peak of the F2 event.  The continuum flux ratios were adequately
explained by the F13 multithread model with decay phase flux
without a hotspot contribution, which is the same scenario for the model for the
decay phase spectrum S\#24 from 
the Megaflare.  When significant time has passed after the peaks of energetic
secondary flares, $X_{hotspot} \sim 0$, such as for the YZ CMi Megaflare at the time of S\#24 and for the DG CVn
Superflare F2 event at $\approx T0+11750$~s and $\approx T0+17,000$~s \citep{Osten2016}.  In future work,
we will constrain $X_{hotspot}(t)$ using a
superposition of sequentially heated hotspot spectra, as done to produce the
F13 multithread model.

In solar flares, there is also evidence that the characteristics of
the beam heating can significantly change between the 
main flare peak and secondary flares.  \citet{Warmuth2009}
describe a secondary flare that produced a hard X-ray spectrum that could be
explained by an electron beam with a higher, low-energy cutoff 
($E_c\sim110$ keV) than in the initial impulsive phase.   The
parameters of the 
electron beam for this flare are less extreme but
qualitatively similar to the parameters for the $E_c=500$ keV electron
beam that
we use to produce a hotspot spectrum.  Our modeling of the hotspot
implies that the heating may result from ultrarelativistic electrons.
Synchrotron emission from ultrarelativistic ($E>500$ keV) electrons
are a possible source for the sub-mm/THz radiation in solar flares \citep[see the review
in][]{Krucker2013}.  Sub-THz radiation attributed to synchrotron emission has also been detected from
stellar flares in active binary systems \citep{Massi2006}.  



\begin{figure}[htbp]
\begin{center}
\includegraphics[scale=0.8]{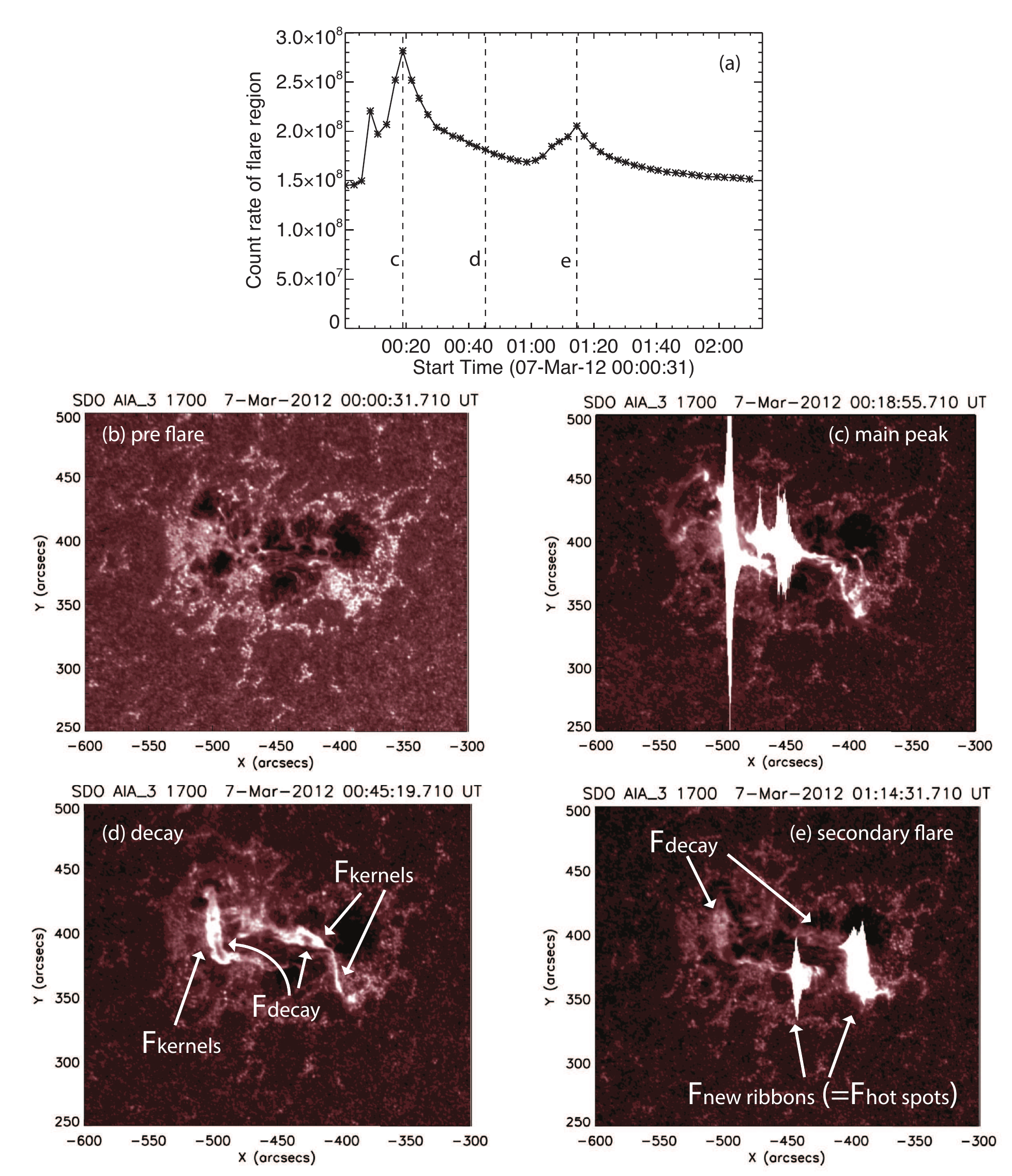}
\caption{(a) Spatially integrated SDO/AIA 1700 light curve of the X-class flares on
  2012 March 7 (GOES classes X5.4 and X1.3 peaking at 00:24 and
  1:14, respectively).   Panels (b)-(e) show the spatial evolution of the intensity
  at representative times indicated by the vertical dashed lines in
  (a). Panel (c) shows that two bright ribbons develop and result in
  brightest intensity during the main
  peak that saturate the image. Panel (d)
shows the flare region as the light curve decays and the ribbons
separate; new kernels develop as these ribbons separate from one another. As the western
ribbon spreads towards the plage near the main umbra, another
two-ribbon event commences; the peak of this event is shown in panel (e).  The separating ribbons in (d) have
decayed in brightness in (e).  We interpret the three model component
surface flux spectra (Figure \ref{fig:megaflare_models}) in the
YZ CMi Megaflare starting $\sim$70 minutes after the flare start using
the spatial morphology in this solar flare:  1) separating ribbons consisting of newly heated
flare kernels ($F_{kernels}$) at the leading edge of the ribbons, 2)
decaying kernels ($F_{decay}$) from each large event in the
wake of the newly heated kernels, 
and 3) a new series of two ribbons that develop ($F_{new\ ribbons}$). In the Megaflare
secondary flare MDSF2, we infer that the heating occurs at high column
mass in the
kernels in these new ribbons, which we refer to as hotspots 
($F_{hotspot}$).  Several potentially analogous features in the development of the 2012
March 7 flare are indicated.  Our model for S\#24 in the Megaflare
suggests a spatial development as in panel (d), and our model for S\#113 suggests a
spatial development as in panel (e) but with kernels still
continuing to develop either in the new ribbons or at the leading edge
of the decaying ribbons.  At the approximate location of the secondary
flare (panel e), there was flare activity at the time of the main flare
peak.  The same image scaling is
used in panels (c)-(d), and $aia\_prep$ was used to generate these data.   }
\label{fig:tworibbons}
\end{center}
\end{figure}

In Figure \ref{fig:tworibbons}(b-e) we show representative images from
the Atmospheric Imaging Assembly \citep[AIA;][]{Lemen2012} on the Solar
Dynamics Observatory (SDO) 
in the 1700 filter during the GOES X5.4 and X1.3 solar flares of 2012
March 7. The AIA 1700 filter shows the footpoint features (ribbons and
kernels) in solar
flares at high spatial resolution.  There are no simultaneous spectral
observations over the wavelength range of this filter, but studies of other solar flares
indicate a combination of continuum, line intensity and
pseudo-continuum intensity from blended lines \citep{Cook1979, Doyle1992,
  Brekke1997, Qiu2013}.  The FUV continuum becomes
bright in stellar flares \citep{HP91}, and it may be a good tracer of the
formation of hot spots in secondary flares \citep{Ayres2015}.
The solar flare event of 2012 March 7 provides a unique comparison to the Megaflare because it
produced two significant flares in the 1700 light curve (Figure \ref{fig:tworibbons}(a)), analogous to a ``big first
flare'' and a high-energy secondary flare, within the same active
region.   Notably, the thermal response
detected by GOES shows a significant difference compared to the secondary flare
in \citet{Warmuth2009} which does not respond in GOES, 
suggesting that the heating mechanism is not uniform among secondary flare
events. 

 In Figure \ref{fig:tworibbons}(b)-(e), we use the spatial development
 in the 2012 March 7th solar flare to illustrate
possible analogous development of the three continuum-emitting
components ($F_{kernels}, F_{decay}, F_{hotspot}$) in the YZ CMi Megaflare. 
 Combining the spatial development in Figure
\ref{fig:tworibbons} with the RHD modeling of the Megaflare decay
phase spectra, a possible scenario for the spatial development of the Megaflare decay phase is as
follows: 1) at the start of the spectral observations, bright kernels with
a filling factor of 0.1\% are heated by high flux electron beams at the leading edges of the
spreading ribbons (panel d);  2) each kernel rapidly
decreases in brightness as beam heating ends.  The total flare area
that has been swept out by these kernels and other ribbons earlier in
the flare has a
filling factor of $\sim1-2.5$\%  and
decays in brightness on a longer
timescale than the rapid brightness decay of each kernel (panel d); 3) later, the total
area of the kernels at the leading edges of the separating flare
ribbons has
decreased in size to a filling factor of 0.05\%.  These kernels 
 propagate into
the locations of the active 
region where new ribbons with hotspots heated by ultrarelativistic
electron beams are
triggered and develop with a filling factor of 0.1\% (panel e).  Also at
this time, the locations of the kernels formed in panel (d) are still
decaying on a long timescale.

\clearpage

\section{Discussion and Future Work} \label{sec:discussion}
The TB09$+$HM88 broadening prescription 
presents a challenge for explaining the white-light continuum flux distribution in
dMe flares using high flux (F13) nonthermal electron beams that generate
very dense chromospheric compressions and much broader lines
than observed.  A multithread modeling approach with the new line broadening gives
a more direct comparison to spatially averaged observations of stellar
flares, which are a spatial superposition of many flare loops at
different stages in their heating and cooling evolution.  
The multithread modeling shows that the average broadening over the time-evolution of the
F13 model is $\sim2$x lower than the instantaneous maximum broadening, but the H$\alpha$ and
H$\beta$ line flux ratios (relative to H$\gamma$) are too small
compared to the 
observations, although the H$\delta$ decrements are in agreement. 
The F13 multithread models are consistent with the observed
continuum flux ratios, line-to-continuum flux ratios, line
broadening, and H$\delta$/H$\gamma$ flux ratio of the early impulsive rise phase but not the main
peak of impulsive-type dMe flare events \citep{Kowalski2013, Kowalski2015}.

We applied a multithread model with several flux components and the TB09$+$HM88 broadening prescription to the phenomenological
model of \citet{Kowalski2012SoPh} for a representative YZ CMi Megaflare decay phase
spectrum.  The observed gradual phase spectrum is a superposition of 
flux components spanning a large range of electron densities and
hydrogen Balmer line broadening.   Using the
multithread modeling with several flux components that was used to explain the white-light
continuum properties in the gradual phase of the DG CVn Superflare
in \citet{Osten2016}, we find that the Balmer
decrements, broadening, and continuum flux ratios are well-reproduced 
in the decay phase spectra of the Megaflare.   In our phenomenological RHD model of the
Megaflare decay phase,
chromospheric condensations develop from high flux electron beam heating 
in new kernels in the leading edge of spreading flare ribbons in the
decay phase of the spatially integrated flux.  
The filling factors of the visible stellar hemisphere for each flux component 
are similar between the RHD models and the previous phenomenological
models. 

During the secondary flare of the Megaflare, a hot spot with a
Vega-like flux component is formed
by relativistic electron beam heating in our RHD,
three-component multithread 
model, which accounts for the observed changes in line broadening,
line decrements, and continuum flux ratios 
through the
secondary flare.  If fully
relativistic electron beams are present in these types of secondary
flares, (triggered) ALMA observations may be able to detect the
synchrotron radiation.   Adding a hotspot spectrum to the spatially superposed
model of the in impulsive-type dMe flares may also explain the 
spectrum continuum flux ratios and Balmer decrement in the main peak of these flares.  Observations for the main peak in the
Megaflare are not available, but these models can be applied to other large events with spectral coverage, such as the 
YZ CMi ``Ultraflare'' \citep{Kowalski2016} or the IF3 event from
\citet{Kowalski2013}.  The model spectra ($F_{kernels}, F_{decay}, F_{hotspot}$) will
be available online for future modeling work. 

An alternative explanation for the hotspot may be heating from MeV
proton beams that are preferentially accelerated in the secondary
flares.  MeV protons are expected to penetrate deeper than deka-keV electrons \citep{Zharkova2007}
and are also thought to be preferentially accelerated in the later
phases of a flare when the reconnected loops have a lower magnetic
field strengths \citep{Petrosian2004}.  Gamma-rays are produced from the
interaction of accelerated protons, alpha particles, and ions with the solar
atmosphere \citep{Vilmer2011}.  Gamma-ray imaging of a gradual phase of
a large solar flare has shown that the
protons are preferentially accelerated in different loops than the
bulk of the electrons \citep{Hurford2006}.  The secondary flare peaking
at 1:15 UT in Figure \ref{fig:tworibbons}
produced hard X-ray and gamma rays detected by Fermi/GBM and the Fermi/LAT,
and the flux detected by the latter was interpreted as evidence
of high energy protons interacting with the lower atmosphere \citep{Ajello2014}. 
 Gamma-rays are too faint to observe from
other stars, but there has been a claimed detection of charge
exchange from proton beams during a flare in the dMe star AU Mic
\citep{Woodgate1992}.  Nonthermal protons and ions are produced in solar flares with
comparable energies to the nonthermal electrons \citep{Emslie2012},
and further work should be dedicated to
understanding the heating from high energy protons in solar and stellar flares,
and in the (possible) role of neutralizing the F13 electron beams that we
use to produce dense, chromospheric condensations.  Without charge
neutralization, 
large return current electric fields and beam instabilities would
limit the propagation of F13 beams.

\subsection{Speculation on the Origin of the Extended Gradual Decay
  Phase Component} \label{sec:gradual}

The H$\alpha$/H$\gamma$ and
H$\beta$/H$\gamma$ decrements are not well reproduced by the
0-5~s average (multithread model) using the F13 
because a high optical depth over the continuum emitting regions leads
to an optical depth in these lower order lines that is very large.
A larger filling factor of the  F13 flux at $t=4$~s (after the beam
heating ends) compared to the impulsively heated kernels is required to produce a better match to the
line flux ratios.   

The large filling factor (1-2\% of the visible stellar hemisphere in the Megaflare) suggests that this
missing component is due to the decaying flux from previous,
larger events (e.g., the main event, or the secondary flare just prior
to the spectral observations).  However,  the timescale
of the flux decay in the models ($\sim$seconds) is far too short to explain
the gradual decay of flux occurring $\sim70$ minutes after the main peak of the Megaflare.
Longer timescales after
beam heating ends in any given kernel are needed to reproduce
the timescales of the extended gradual decay phase white-light
emission component in
megaflares and in lower energy, classical flares
\citep{Davenport2014}. 

Several heating mechanisms have not yet been critically tested with
RHD models and could have an important role in producing extended gradual
decay phase flux. 
We speculate that 3D backwarming from X-rays and/or UV/EUV (300-3000 \AA)
line emission \citep[``the metal line backheating
hypothesis'';][]{Machado1989, HF92, Fisher2012} will help to account for the missing
flux in Ca II K line, the flux in the Balmer lines, and the Balmer
decrement for the models.  The models of
\citet{HF92} have shown that irradiation from X-rays can heat the
chromosphere and produce a large ratio of emergent line-to-continuum
flux.  The Fe II flux
in the NUV becomes bright in dMe flares \citep{HP91,
  Hawley2007} and may also contribute significantly to the backwarming.  
 Three-dimensional backwarming from UV/EUV lines
has recently been
discussed as a source of the slowly decaying component in optical flare kernels
\citep{Kawate2016}.  Extending the results from 
1D modeling of backwarming from Balmer continuum photons
\citep{Allred2006} to 3D may also explain the
observations \citep{Metcalf1990}.   

After a period of fast retraction from magnetic reconnection,
flare loops are expected to continue to slowly contract \citep{Longcope2011}; betatron
acceleration of particles in these loops may also contribute as a 
heating source in the extended gradual phase.  
Alfven waves are expected to carry Poynting flux and heat the
chromosphere with a delay from reconnection that is inversely
proportional to magnetic field strength
\citep{Fletcher2008, Reep2016}.  Our line broadening prescription will be incorporated
into the publicly available version of the RH code and will be available to test
the line decrement and broadening predictions from
radiative-hydrodynamic multithread modeling of high-flux density electron (and
proton/ion) beam heating combined with each of these additional
heating scenarios.  The effects of non-thermal
collision rates on the Balmer line wings \citep{Canfield1987,
  Kasparova2009} may also be revisited with the new
prescription for electric pressure broadening in order to constrain the beam fluxes in
the newly formed kernels.  Modeling of the red-wing asymmetry in
solar flares \citep{Ichimoto1984, Sharykin2014} will also benefit by the
new broadening modeling of the hydrogen wings.

\section{Summary \& Conclusions}\label{sec:conc}

Using the TB09$+$HM88 line profiles, we have incorporated the VCS theory of electron/proton pressure broadening
into the RH code for accurate modeling of the broadening of the
hydrogen lines during flares.  This resolves the $\sim$ order-of-magnitude
discrepancy in the inferred ambient charge density from Balmer line
broadening discussed extensively by \citet{1997ApJS..112..221J}.  
Convolving the Voigt profile
function with the theoretical profiles of TB09$+$HM88 extended from the VCS unified
theory produces much broader Balmer
lines than using a Voigt profile with an electric pressure damping parameter from 
the analytic approximations of S78.  The
wavelength-integrated line fluxes of Balmer lines can be several
times larger with the TB09$+$HM88 profiles.  The $\lambda>3700$ \AA\ spectrum 
calculated with our method produces a spectrum of Vega that is much
more consistent with the observations of the broadening and higher
order line merging than with the S78 prescription previously used in
RH.  Combined with the opacity effects from level dissolution at the
Balmer edge, the new modeling prescription
with the theoretical TB09$+$HM88 profiles provides self-consistent,
robust constraints on flare heating
model predictions for the flare-enhanced charge density in the lower
atmosphere.  
Identifying the highest order, undissolved Balmer line resolves ambiguity 
in the charge density inferred from the broadening in the far wings of lower order Balmer
lines, which is model dependent.  Although our method for calculating
the opacity effects from level dissolution is sufficient for our
purposes here, we re-iterate the suggestion of
\citet{Kowalski2015} to include the occupational
probability formalism of \citet{HHL} in the non-equilibrium rate
equation as new flare codes (or significant improvements to existing
flare codes) are developed in the future.

We revisited the phenomenological model of the YZ CMi Megaflare with the
new broadening prescription and new RHD models.  A superposition of three emitting regions is
necessary to reproduce the general range of values and the evolution of the Balmer
decrements, broadening, and continuum flux ratios:  

\begin{enumerate}
\item a flux component (filling factor of 0.1\%) from downward-directed, heated compressions (condensations) that result from high flux
density electron beams heating the upper chromosphere in many
simultaneously heated and cooling flare loops (threads); 

\item a flux component (filling factor of 0.1\%) from
ultra-relativistic electron beam heating
at high column mass (near the
lower pre-flare chromosphere); and

\item  a flux component (filling
factor of 1-2\%) with bright, narrow line
flux that decays on a longer timescale than currently produced in
the models. 

\end{enumerate}

 The flux components from \#1 and \#2 result from atmospheric heating to
 $T=12,000-13,000$ K
at high column mass $\sim0.01-0.05$ g cm$^{-2}$, at heights from
$z\sim150-250$ km (corresponding to the lower chromosphere in the 
pre-flare atmosphere). These two heating simulations largely differ in
the hydrodynamics; most
of the continuum intensity forms over a narrow (several km) region in
the chromospheric condensation in heating scenario \#1 \citep{Kowalski2015}, whereas the
continuum intensity in heating scenario \#2 forms over a 100 km
uncompressed region at high column mass.  The 
  Balmer line radiation in the emergent flux spectra forms over large optical
  depth and high electron density in these new
  RHD flare models.  Correct modeling of these profiles
  therefore requires an
   accurate treatment of the far wing broadening, such as given by the VCS theory as
   implemented in \citet{Tremblay2009}. 
The charge density values that produce the broadening the Balmer lines
in the superposed model spectra span a large
range of
$n_{e} \sim 10^{14}-5\times10^{15}$ cm$^{-3}$; the densities that
broaden a Balmer line vary as a function
of wavelength due to the optical 
depth variation over a line and due to the time-evolution of $n_e$ in
each model.   
In the superposed flux spectra models of the YZ CMi Megaflare, we require more emission line
flux in the Ca II K and Balmer lines, which is
evidence that flux component \#3 needs improvement or that 
an additional component is required to produce heating over an area of the star with lower charge
density than \#3; one possible heating source is 3D
radiative backheating.

In solar flares, the broadening evolution on several second
timescales in spatially resolved kernels can be
constrained with future spectral observations with the
Daniel K. Inouye Solar Telescope.  As shown from the modeling of the YZ CMi
Megaflare, 
spatially resolved spectra of the hydrogen
lines and hydrogen edge wavelength regions during large
solar events would place strong constraints on the variation of the
charge density and heating mechanisms across 
flare ribbons.



\section*{Appendix A:  High $E_c$ RHD Flare Models}

As discussed in \citet{Kowalski2016}, we are working on a large grid
of RADYN flare models where we vary the parameters of the nonthermal
electron beam and have found that models with a high value of $E_c$ 
cause the beam energy to penetrate deeper than values near $20-37$
keV, which are typical values in solar flares.  High $E_c$ models also
contain
fewer electrons for the same energy flux as lower $E_c$ models, and
thus
there are smaller effects from the return current and beam
instabilities.  
 Here we describe two representative
models from our grid of dMe flare models that reproduce key
properties of dMe white-light continuum flux and have important differences
in the broadening and decrements of flare spectra than lower $E_c$ models.  

\subsection*{Appendix A.1:  The Starting Atmosphere}

We update the starting dMe pre-flare atmosphere from
\citet{Kowalski2015} using the prescription for incident
XUEV radiation described in \citet{Allred2015}.  Charge conservation
is calculated using the ionization from the detailed elements hydrogen
and helium, but for calcium the electron contribution to charge
conservation is calculated from LTE (as for the remaining elements).  
In future work, we intend to include a Ca I-II-III model atom in RADYN 
which is appropriate for the NLTE ionization and charge contribution
for an M dwarf atmosphere (in the Sun, Ca II is dominant and a calcium
ion without the neutral stage
is sufficient for the detailed losses in RADYN).  For atoms and ions not treated in detail,  we use
the radiative loss function from \citet{Allred2015} which includes the most updated losses
from CHIANTI at $T>20,000$ K but excludes several ions
at low temperature at $T<20,000$ K that are optically thick in flares (e.g., Fe II, Si
II, Mg II, Al II) as discussed by \citet{RC83} and \citet{HF92}.  We
converge the atmosphere top boundary that is reflecting with
a zero temperature gradient.  We also apply a small
amount of non-radiative heating (0.23 erg cm$^{-3}$ s$^{-1}$) to all
heights with $T>1$ MK for a temperature gradient in the
corona that produces a transition from conductive cooling to
conductive heating at the transition region \citep{Klimchuk2008}.
The atmosphere is relaxed with this condition and this heating is
applied through the flare simulations to keep the corona from cooling
below the pre-flare temperature.  
We also note that the chemical equilibrium of hydrogen and H$_2$ is
included using LTE dissociation, which is important in the
photosphere.  Continuum wavelengths that have been added to the detailed
radiative transfer are $\lambda=$1332, 1358, 1389, 1407, 1435, 1610, 2519, 2671,
2780, 2826, 3300, 3500, 3615, 4170, and 6010 \AA.  These are needed
for direct comparisons to data.

\subsection{Appendix A.2:  The $E_c=500$ keV Flare Model}
We apply electron beam heating for 2.3~s with $E_c=500$ keV,
$\delta=7$, and an energy flux density of $2\times10^{12}$ erg
cm$^{-2}$ s$^{-1}$ (2F12).  The atmosphere is allowed to relax for 60
seconds after 2.3~s.  These parameters were chosen so that the beam
energy is localized to the lower atmosphere and results in a local
temperature maximum near $T\sim10,000$ K within a short time to be
consistent with the short pulses observed during solar and dMe
flares \citep{Aschwanden1995, Robinson1995}.  

\begin{figure}[htbp!]
\plotone{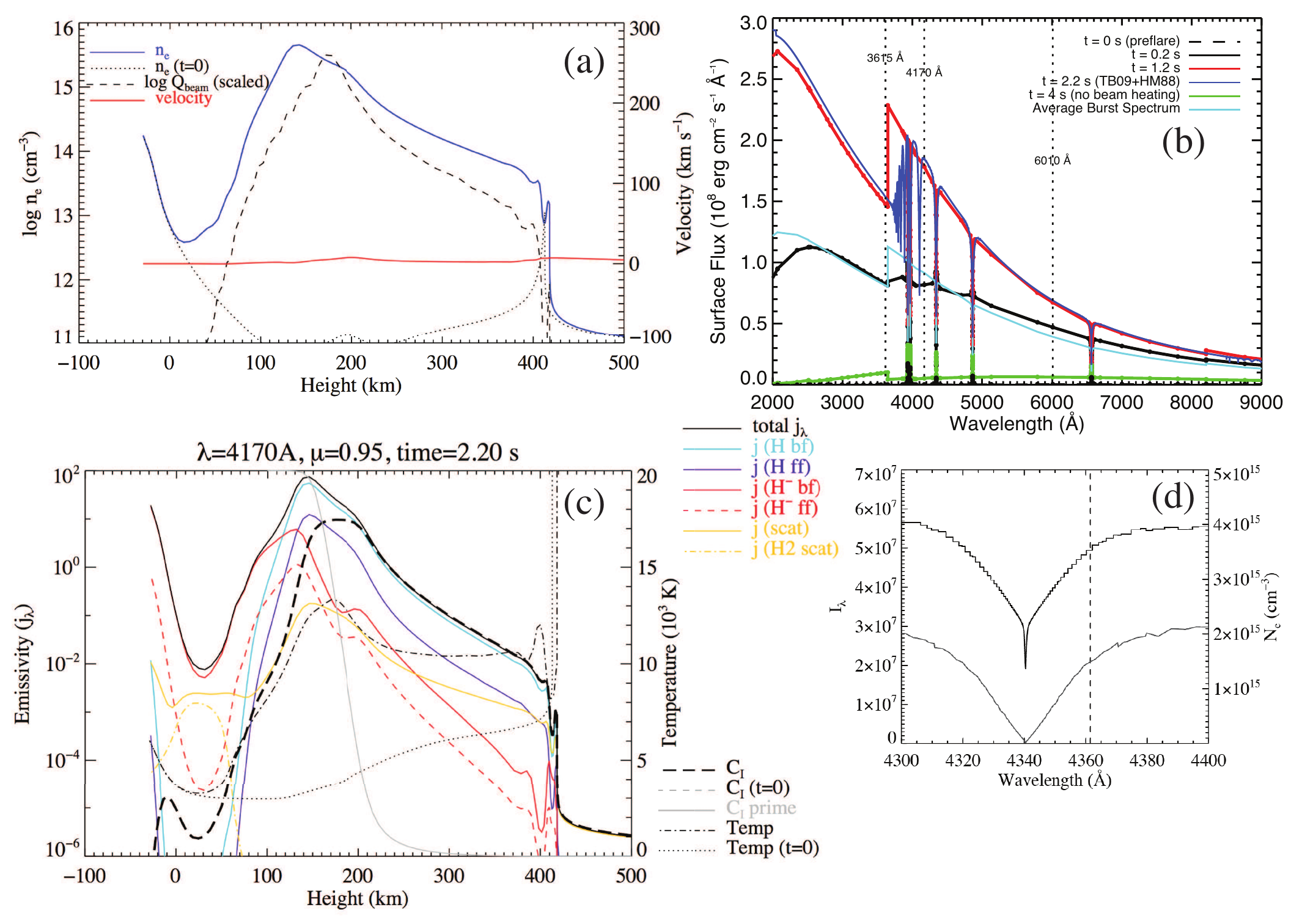}
\caption{(a) Electron density vs. height at $t=2.2$~s in the $E_c=500$
  keV heating model.  The volumetric beam heating ($Q_{\rm{beam}}$; erg cm$^{-3}$
  s$^{-1}$) is shown scaled to the left axis over 5 dex.   (b) The
  evolution of the detailed  flux
  spectrum from RADYN at $t=0, 0.2, 1.2$ and 4~s; the spectrum from RH
  at $t=2.2$~s is shown with the TB09$+$HM88 broadening and the opacity
  from level dissolution. (c)  Sources of continuum emissivity that contribute to the emergent
  continuum intensity at $\lambda=4170$ \AA\ at $t=2.2$~s in the
  $E_c=500$ keV heating model.  The contribution function (thick
  dashed line) to the
  emergent intensity and the cumulative contribution function (gray; $C_I^{\prime}$)
  show that the continuum is formed over a ``hot spot'' with a peak
  temperature of nearly 13,000 K.
  The dominant source of emissivity over the hot spot is spontaneous hydrogen
  recombination emissivity. (d) The emergent intensity at $\mu=0.95$
  for the
  H$\gamma$ line with the TB09$+$HM88 broadening (black line);  the contribution-function
  weighted electron density is shown on the right axis (gray line), and a vertical
  dashed line indicates the same wavelength in the far red wing in Figure
  \ref{fig:vegacf} (bottom).
\label{fig:ec500}}
\end{figure}

In Figure \ref{fig:ec500} we show the physical parameters at
$t=2.2$~s in the lower atmosphere: the
electron density variation, gas velocity, and beam heating in panel
(a) the emergent flux spectra at representative times ($F_{hotspot}(t)$; see text) in panel (b), the contribution function to the
emergent continuum intensity and the continuum emissivity processes at $\lambda=4170$ \AA\ in panel (c) using the method described in
\citet{Kowalski2016IRIS}, and an inset of the H$\gamma$
line with the TB09$+$HM88 broadening calculated from RH in panel (d).  The gray line in panel (c) shows the cumulative contribution
function ($C_I^{\prime}$) normalized from 0 to 1 on the right axis, which illustrates the depth range over which the emergent
intensity is formed.  The blue ($\lambda=4170$ \AA) continuum forms over $z=150-235$ km
which corresponds to a log column mass range of log $m$/g
cm$^{-2}=-1.26$ to $-2.05$, a
large range of 
electron density from $5-40\times10^{14}$ cm$^{-3}$ (with a
contribution-function weighted average of $2.3\times10^{15}$ cm$^{-3}$), and a temperature
range from $T=10,800-12,700$ K.   The electron density weighted by the
contribution function for the emergent intensity over the H$\gamma$
line ranges from $2\times10^{13}$ cm$^{-3}$ at line center to
$1.8\times10^{15}$ cm$^{-3}$ in the far wing ($\lambda_{\rm{rest}}+20$ \AA).  The broadening is
comparable to the Vega spectrum of H$\gamma$, which forms over an
electron density range that is significantly lower ($n_e=10^{12}$
cm$^{-3}$ at line center and only $5\times10^{14}$ cm$^{-3}$ in the far
wings).  Vega has a factor of nearly 100 lower gravity compared to the
M dwarf, and thus the radiation is formed over $\sim1000-2000$ km in
the line wing (Figure \ref{fig:vegacf}) compared to 85 km in the line
wing of the M dwarf $E_c=500$ keV model. Thus, a comparable opacity is
  produced over a larger range of heights with lower density in Vega.
In the Balmer edge region, the highest order undissolved Balmer line
in absorption is $\approx$H13 in the $E_c=500$ keV model (Figure \ref{fig:megaflare_models}),
whereas in the Vega model (Figure \ref{fig:vega} top) it is H16 or
higher, as expected from the density dependence on the occupational
probability of higher levels \citep[cf Figure 9 of][]{Kowalski2015}.  

 Using the equations in \citet{Holman2012} for the energy lost due to
the return current electric field (assuming that proton beams do not
neutralize the electron beam), the energy lost over the top 8 Mm of
the corona is 4 keV / electron and the Joule heating rate is $\sim15$ erg
cm$^{-3}$ s$^{-1}$.  The 500 keV electrons lose a small fraction
of their initial energy and thus we don't expect the beam spectrum to
be modified significantly.  The coronal heating rate from the return
current is 65x the heating rate necessary to keep the pre-flare corona
hot.  This compares to nearly 8000 erg cm$^{-3}$ s$^{-1}$ that results
from an F13 beam with $E_c=37$ keV and $\delta=3$.  The return current
drift speed is $<$ 10\% of the electron thermal speed in the corona,
and therefore we do not expect energy-draining double-layers to
develop \citep{Lee2008, Li2014}; for the F13 beam the drift speed
is three times the electron thermal speed and double layers will form.

\subsection{Appendix A.3:  The $E_c=150$ keV Flare Model}

We apply electron beam heating for 2.3~s with $E_c=150$ keV,
$\delta=3$, and an energy flux density of $5\times10^{12}$ erg
cm$^{-2}$ s$^{-1}$ (5F12). The atmosphere is allowed to relax for 60
seconds after 2.3~s.  For this beam, we calculate that $<20$ keV is lost per electron in the
propagation over the top 8 Mm of the loop and there is a Joule heating rate from
the return current of 350  erg cm$^{-3}$ s$^{-1}$ which will result
in a temperature change.  The drift speed of the return
current is 40\% of the electron thermal speed in the corona, and thus we do not expect energy-draining double-layers to
develop \citep{Lee2008, Li2014}.  

The model results are shown in
Figure \ref{fig:ec150} for the same panels as for the $E_c=500$ keV
model in Appendix A.2.   Compared to the phenomenological models of \citet{Cram1982}, 
the temperature profile, density, and emergent continuum spectrum are
similar to the extreme model \#5, whereas the broad
Balmer wings with deep central reversals are similar to the emergent
H$\alpha$ profile from model \#3 that exhibits a low column
mass transition region.  In the $E_c=150$ keV model the transition
region is at low column mass as in the \citet{Cram1982} model \#3, but the
temperature and electron density values at high column mass (log
$m$/g cm$^{-2}=-2$) are larger than the \citet{Cram1982} model \#5.  

Compared to the 2F12, $E_c=500$, $\delta=7$
heating model (Appendix A.2), the blue ($\lambda=4170$ \AA) continuum forms
over comparable heights ($z\sim150-250$ km; see the $C_I^{\prime}$ curves), column masses (log $m$/g
cm$^{-2} > - 2.2$), temperatures ($\sim12,000-13,000 $ K), and electron
densities ($n_e=5-40 \times10^{14}$ cm$^{-3}$ with the same contribution
function-weighted electron density of  $2.3\times10^{15}$ cm$^{-3}$).  The relative importance
of each emissivity process in the formation of the continuum intensity
(panel c in Figures \ref{fig:ec500} and \ref{fig:ec150})
is very similar for the two models, while the contribution
function-weighted electron density over the H$\gamma$ lines (panel d
in Figures \ref{fig:ec500} and \ref{fig:ec150}) is also 
similar.   In both heating simulations, the formation of the emergent continuum intensity at
$\lambda=3615$ \AA\ is shifted 50 km higher than the formation of the 
$\lambda=4170$ \AA\ emergent continuum intensity because of the larger 
Balmer ($n=2\rightarrow\infty$) bound-free opacity compared to the
Paschen ($n=3\rightarrow\infty$) bound-free opacity.  As a result,
the $\lambda=3615$ \AA\ 
emergent continuum intensity originates from a lower electron density than
the $\lambda=4170$ \AA\ continuum intensity.   This produces much smaller
Balmer jump ratios ($F_{3615}/F_{4170}$) in the model spectra than for 
a spectrum that results from hydrogen recombination over low optical
depth.  In contrast, the F13 model spectra exhibit small Balmer
jump ratios because of the large variation of the physical depth range
as a function of wavelength, as described in \citet{Kowalski2015,
  Kowalski2015IAU, Kowalski2016}.  The F13 model also has two flaring
layers with an electron density profile that increases outward, in
contrast to the high $E_c$ models which exhibit electron density
profiles that increase towards lower heights.  The electron density
variation over the two flaring layers combined with the optical depth 
variation (resulting in the physical depth range variation) as a function of wavelength both contribute to the
characteristics 
of the emergent flux spectrum in the F13 model.   

The striking differences in the emergent flux and intensity spectra
of the Balmer lines and at wavelengths in the Balmer continuum for
the $E_c=150$ and $E_c=500$ keV models can
be explained with the panels in Figures \ref{fig:ec500} and \ref{fig:ec150}.  The lower energy
cutoff ($E_c=150 $ keV) 
heats the mid and upper chromosphere to a much larger extent than the
$E_c=500$ keV model;  the beam heating peaks at $z\sim280$ km in the former 
and at $z\sim180$ km in the latter (panel a of Figures \ref{fig:ec500}
and \ref{fig:ec150}).  The larger heating rates at $z>250$ km in the
$E_c=150$ keV model 
produce a larger ionization fraction and excitation at these heights, which leads to a
lower optical depth in the Balmer continuum and Balmer lines.  For
example, $\tau_{3615} = 1$ occurs at $z=210$ km in the $E_c=150$ keV model and at
$z=225$ km in the $E_c=500$ keV model.  
For the emergent Balmer continuum flux (e.g., $\lambda = 3615$ \AA) in the
$E_c=150$ keV model, the
photons are formed deeper and at higher $n_e$: at $\tau_{3615}=1$ 
the electron density ($1.3\times10^{15}$ cm$^{-3}$) is 1.8x larger
than the density ($7\times10^{14}$ cm$^{-3}$) at $\tau_{3615}=1$ in the
$E_c=500$ keV model, which results in a larger
continuum emissivity ($\propto n_e^2$) at the heights where $\tau_{3615}=1$ and a
larger emergent continuum flux.  Therefore, the emergent Balmer continuum flux in
the $E_c=150$ keV model is relatively brighter than the
$\lambda=4170$ \AA\ flux  (the Balmer continuum flux is ``in
emission'')  whereas the
Balmer continuum flux in the $E_c=500$ keV is fainter
than the $\lambda=4170$ \AA\ continuum flux (the Balmer continuum
flux is ``in absorption'').  The formation of
the  $\lambda=4170$ \AA\ continuum flux is similar in the two models because
the $E_c=150$ keV model has a very hard distribution ($\delta=3$) and a
higher flux which heats the deeper layers to a comparable level as the softer,
lower flux, higher-energy $E_c=500$ keV model. 

 At $\lambda_{\rm{rest}}+2.5$
\AA, the $E_c=150$ keV model H$\gamma$ intensity is in emission while the
$E_c=500$ keV model H$\gamma$ intensity is in absorption.  Although
the electron density for the line formation appears similar in panel (d) of
Figures \ref{fig:ec500} and \ref{fig:ec150}, there is a factor of 1.5x
larger electron density for the $E_c=150$ keV model at $\lambda_{\rm{rest}}+2.5$
\AA.  Because of the
higher temperature at $z>200$ km, the optical depth is lower in
the line and H$\gamma$ is formed deeper where there is higher electron
density.  The higher temperature over the line formation for the
$E_c=150$ keV model also
contributes to a larger emergent intensity because $n_u$ for H$\gamma$
has a larger
population density due to the scaling of the collisional rates with temperature via the Boltzmann exponential factor ($n_u$ and $n_l$ for H$\gamma$ are near LTE at the height of the 
maximum in the contribution function in these models).

\begin{figure}[htbp!]
\plotone{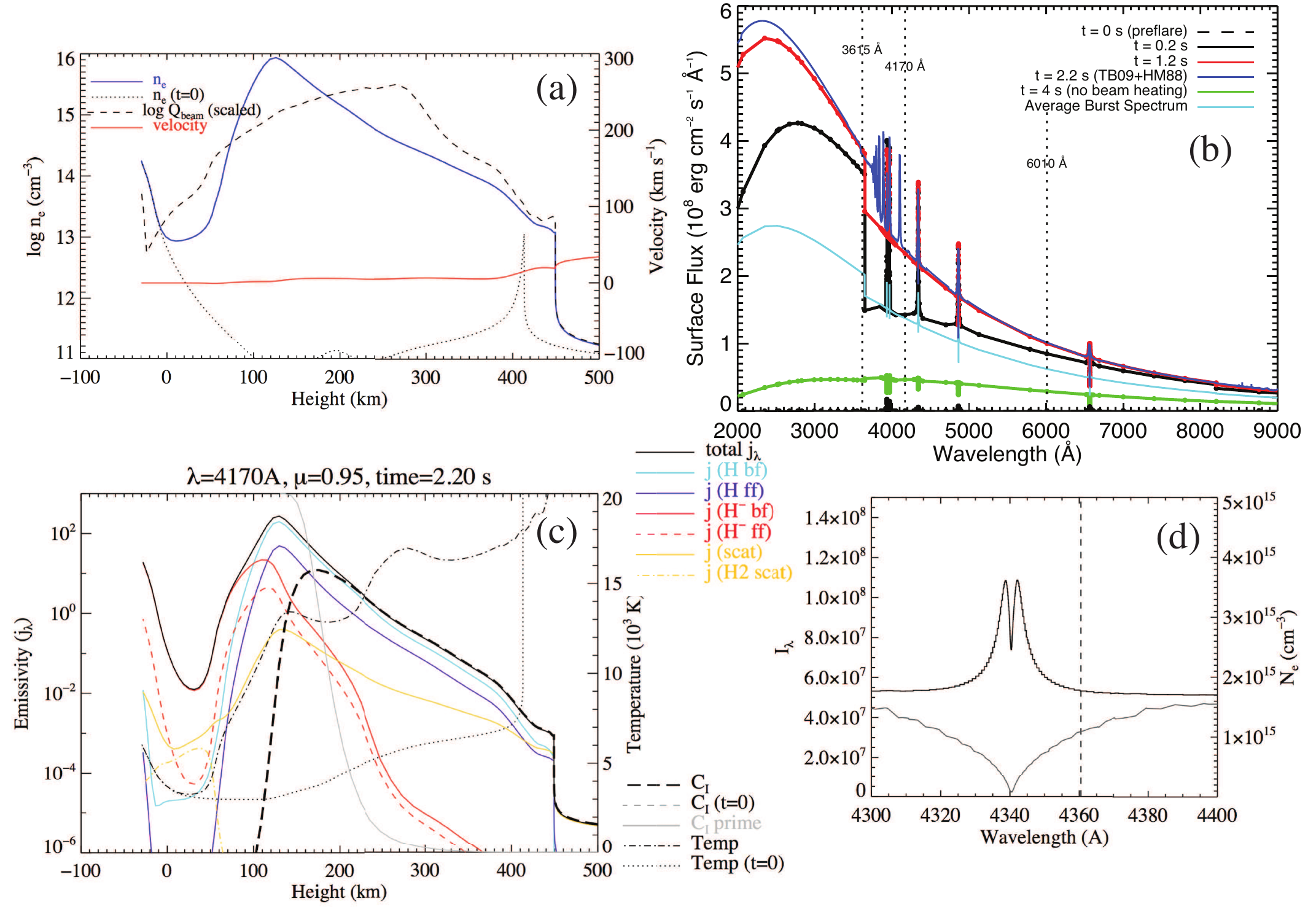}
\caption{We show the same panels as in Figure \ref{fig:ec500} for the
  $E_c=150$ keV, $\delta=3$, 5F12 electron beam heating model.  }
\label{fig:ec150}
\end{figure}

\clearpage
\section*{Appendix B:  Terminology \& Abbreviations}

In this appendix, we show a list of abbreviations and terminology
used throughout the text.  

\begin{itemize}

\item \textbf{MDSF2}: ``Megaflare Decay Secondary Flare \#2''
  (see Figure \ref{fig:megaflare_lc}).

\item \textbf{S\#24}:  Spectrum \# 24 in the decay phase of the YZ CMi
  Megaflare (see Figure \ref{fig:megaflare_lc}).

\item \textbf{S\#113}:  Spectrum \# 113 at the peak of the MDSF2 event in
  the YZ CMi Megaflare (see Figure \ref{fig:megaflare_lc}).

\item \textbf{VCS}:  The unified theory of electron \& proton pressure broadening
presented in \citet{1971JQSRT..11..263V, Vidal1973}.

\item \textbf{TB09$+$HM88}  The profiles calculated with the VCS
  unified theory extended with the modifications of
  \citet{Tremblay2009}, the occupational probability formalism of
  \citet{HM88}, and the non-ideal, pseudo-continuum opacity of \citet{Dappen1987}.

\item \textbf{S78}:  The analytic prescription for electron \& proton pressure
broadening from \citet{Sutton1978} employed in the RH code using a 
damping parameter, $\Gamma_S$, in the Voigt function. 

\item \textbf{$E_c$}: Lowest energy electron in the power-law
  distribution of a nonthermal electron beam.   

\item \textbf{F13}: Nonthermal electron beam energy flux
  density of $10^{13}$ erg cm$^{-2}$ s$^{-1}$ and $E_c=37$ keV.  

\item \textbf{F13 dpl}: F13 heating simulation from \citet{Kowalski2015} with a
double-power law electron beam distribution parameterized by
$\delta=3$ at $E<105$ keV and $\delta=4$ at $E>105$ keV.

\item \textbf{F13 $\delta=3$}: F13 flare simulation from \citet{Kowalski2016} with a
single power-law (spl) electron beam distribution parametrized by a power law
index of $\delta=3$. 

\item \textbf{C3615}: The flare continuum specific flux averaged over $\lambda=3600-3630$ \AA.

\item \textbf{H$\gamma$/C4170}: The line-integrated flare flux divided by the
flare continuum specific flux averaged over $\lambda=4155-4185$ \AA.

\item \textbf{multithread model}:  0-5~s average (0-5~s ave) of an F13
  simulation; also referred to as ``average burst spectrum''.  

\item \textbf{DG CVn Superflare multithread model}:  0-5~s average of
  an F13 model in addition to the instantaneous F13 model at $t=4$~s with a 25 times
  greater filling factor than the 0-5~s average F13 model.  

\item \textbf{pseudo-continuum}:  The non-ideal Balmer continuum opacity 
longward of the Balmer limit wavelength is referred to as the
pseudo-continuum opacity, or dissolved level continuum opacity
(referred to as L-Z Balmer continuum opacity in \citet{Kowalski2015}.  
\end{itemize}




\acknowledgements
We thank an anonymous referee for helpful comments on the manuscript
and for important insights about future directions in flare research using the new broadening modeling. AFK thanks Dr. G. Cauzzi for helpful discussions on flare ribbon development in DST/IBIS data and Dr. D. Graham for 
helpful discussions on flare area calculations.  AFK thanks Drs. L. Fletcher and H. Hudson for
helpful discussions on line broadening while at the University of Glasgow.  
AFK thanks Dr. H. Ludwig for pointing out white-dwarf modeling papers 
at Dr. S. Wedemeyer's workshop at the International Space Science Institute in Bern, Switzerland.
 AFK acknowledges funding that supported this work from the NASA Heliophysics Guest Investigator Grant NNX15AF49G, funding from the University of Maryland Goddard Planetary Heliophysics Institute (GPHI) Task 132, and funding from HST GO 13323. Support for program \#13323 was provided by NASA through a grant from the Space Telescope Science Institute, which is operated by the Association of Universities for Research in Astronomy, Inc., under NASA contract NAS 5-26555.   S. Brown is grateful for the support of STFC Quota Studentship ST/M503502/1.  
\clearpage
\bibliography{starkpaper}

\end{document}